\documentclass{natureprintstyle}

\spacing{1}
\spacing{2}
\spacing{1}

\usepackage{amsmath,amssymb}
\usepackage{graphicx}
\usepackage{color}
\usepackage{bm}
\usepackage{longtable}
\usepackage{amssymb}
\usepackage{rotating}
\usepackage{hyperref}

\newcommand{\aap}{AAP}
\newcommand{\apj}{ApJ}
\newcommand{\mnras}{MNRAS}
\newcommand{\apjs}{ApJS}

\usepackage[utf8]{inputenc}
\usepackage{ulem}
\usepackage{url}
\usepackage{multirow}
\usepackage{xr}
\usepackage{graphicx}
\usepackage{cleveref}
\usepackage{caption}
\usepackage{etoolbox}  
\graphicspath{./supplemental_figures}

\crefname{figure}{Extended Figure}{Extended Figures}
\crefname{table}{Extended Table}{Extended Tables}

\let\citep\cite
\let\citet\cite

\title{An interference-based method for the detection of strongly lensed gravitational waves}

\author{Xikai Shan$^{1,2,3}$, Bin Hu$^{\ast1,2}$, Xuechun Chen$^{1,2}$, Rong-Gen Cai$^{4,5}$}

\begin{document}

\maketitle

\begin{affiliations}
\item Institute for Frontier in Astronomy and Astrophysics, Beijing Normal University, Beijing, 102206, China\\
\item School of Physics and Astronomy, Beijing Normal University, Beijing 100875, China\\
\item Department of Astronomy, Tsinghua University, Beijing 100084, China\\
\item Institute of Fundamental Physics and Quantum Technology, Ningbo University, Ningbo, 315211,  China\\
\item Institute of Theoretical Physics, Chinese Academy of Sciences, Beijing 100190, China\\
\end{affiliations}

\let\thefootnote\relax\footnote{
Corresponding author: Bin Hu: bhu@bnu.edu.cn
}

\vspace{-3.5mm}

\begin{abstract}
\sffamily
The strongly lensed gravitational wave (SLGW) is a promising transient phenomenon.  
However, the long-wave nature of gravitational waves poses a significant challenge in identification of its host galaxy.  
To tackle this challenge, we propose a method triggered by the wave optics effect of microlensing.
The microlensing interference introduce frequency-dependent fluctuations in the waveform.
Our method consists of three steps.
First, we reconstruct the waveforms by using the template-independent and template-dependent methods. The mismatch of two reconstructions serves as an indicator of SLGWs.
This step can identify approximately $10\%$ SLGWs.
Second, we pair the SLGWs' multiple-images by employing the sky localization overlapping.
Because we have pre-identified at least one image through microlensing, the false alarm probability for pairing SLGWs is significantly reduced. 
Third, we search the host galaxy by requiring the consistency of time-delays between galaxy-galaxy lensing and SLGW. 
By combing the stage-IV galaxy survey and the third-generation gravitational wave detectors, we expect to find, on average, 1 quadruple-image system per 3 years. 
The merit of this method can significantly facilitate the pursuit of time-delay cosmography, discovery of compact objects and multi-messenger astronomy.
\end{abstract}

\sffamily


In the past O1-O3 runs~\citep{LIGOScientific:2018mvr, abbott2021gwtc2, LIGOScientific:2021djp}, advanced LIGO~\citep{2015aLIGO}, Virgo~\citep{Acernese_2014}, and KAGRA~\citep{KAGRA:2018plz} (LVK) collaboration has recognized $90$ gravitational wave (GW) events, including $86$ binary black holes (BBHs), $2$ binary neutron stars (BNSs), and $2$ neutron star--black hole binaries (NSBHs).
In the coming years, LVK will continue improving their sensitivity, and LIGO India~\citep{Unnikrishnan:2013qwa} will join the network in the near future.
It is expected that the accumulation of the GW events will rapidly increase with the improvement of detector sensitivity.
Refs.~\citep{Li:2018prc, Oguri:2018muv} predicted that the lensing detection rate for these upgraded second-generation (2G) detectors is $0.5\--1$ per year, consistent with current non-detection~\citep{Hannuksela:2019kle, Kim:2020xkm, McIsaac:2019use, Liu:2020par, LIGOScientific:2021izm, LIGOScientific:2023bwz}.
In contrast, for the third-generation (3G) detectors, such as Einstein Telescope~\citep{2014ASSL..404..333P} and Cosmic Explorer (CE)~\citep{Evans:2016mbw}, the lensing detection rate will increase to $40\--10^{3}~\mathrm{yr}^{-1}$, depending on the population properties of the sources and lenses \citep{Xu:2021bfn}.

The successful detection of SLGW events could have significant implications for both cosmology and astrophysics. In cosmology, SLGWs offer the potential for more accurate Hubble parameter estimation, thanks to the millisecond-level time-delay measurements. Additionally, SLGWs could improve BBH localization precision~\citep{Hannuksela:2020xor, Yu:2020agu} and provide a valuable tool for testing general relativity~\citep{Baker:2016reh, Collett:2016dey,Fan:2016swi}. In astrophysics, the characteristic oscillatory behavior in SLGW waveforms, caused by wave optics effects as the frequency sweeps upward, could spectralize the microlens’s mass distribution, ranging from intermediate-mass black holes to sub-stellar compact objects. This provides a novel approach to studying faint compact objects in galaxies. Unlike electromagnetic signals, the wavelength of GWs during the BBH merger phase is about \( GM \), which is comparable to the size of the source, approximately \( 3GM \). This long wave nature makes the sky localization of GW events much worse than those of the electromagnetic phenomena. In the geometric optics limit, lensing magnification is highly degenerate with the GW's luminosity distance. It is hard to use the magnification information to select the possible lensing candidates.
Therefore, distinguishing lensing events from a vast unlensed dataset is a formidable challenge.
A key issue is to reduce the false alarm probability (FAP).

Four strategies for identification of SLGWs have been proposed in recent years: parameter overlapping~\citep{Haris:2018vmn}, machine learning~\citep{Kim:2020xkm, Goyal:2021hxv}, joint-parameter estimation (joint-PE)~\citep{Lo:2021nae, Liu:2020par, Janquart:2021qov}, and saddle image analysis with high-order modes~\citep{Dai:2017huk, Wang:2021kzt}.
The first two strategies exhibit a comparable FAP~\citep{Goyal:2021hxv}.
They can identify $10\%\--15\%$ lens pairs with a FAP per pair of $10^{-5}$ for 2G detectors.
We extrapolate this detection efficiency to 3G detectors. Assuming that there are $100$ lens pairs and $10^{5}$ unlensed events annually, this method could potentially pick out $10$ to $15$ lens pair candidates along with $50000$ random pairs which are rejected by the null hypothesis (unlensed hypothesis). 

Although we may slightly overestimate FAP, we believe that it is not significantly overestimated. 
The confidence is rooted in the similarity of uncertainties of sky localization between 2G and 3G detectors.
This distribution ranges from $10^{-2}$ degrees to $10^{4}$ degrees~\citep{2017PhRvD..95f4052V}, indicating that many random cases with high coincidences will persist. 
For this reason, Caliskan {\it et al.}~\citep{Caliskan:2022wbh} have argued for the necessity of designing alternative identification criteria beyond the parameter overlapping.
Recently, two possible avenues for such alternatives have been proposed. 
The first involves the incorporation of prior knowledge, including time-delay and magnification ratio between lensing image pairs, as advanced identification criteria. 
The second avenue centers on employing a more accurate joint-PE method to enhance identification capabilities. 
Currently, LVK collaboration has combined this joint-PE method with time-delay priors to determine whether or not an event pair/triplet/quadruplet is strongly lensed\cite{Hannuksela:2019kle,LIGOScientific:2021izm, LIGOScientific:2023bwz}.

However, both the overlapping and joint-PE methods face challenges in the future GW detection missions. 
The computational demands are substantial, with a complexity proportional to $\mathcal{O}(N^2)$, where $N$ represents the number of GW events. 
Therefore, it is necessary to devise a new, prior-free, low FAP, and computationally efficient method as an independent alternative approach to identify SLGWs.
Thanks to the long wave nature of GWs, microlenses (e.g., stars and compact objects) located within the lens galaxies could leave diffraction or interference imprints on GW's waveform, which could be treated as a smoking gun for strong lensing events. 
Our strategy is to leverage these inherent features in SLGWs images.
Ref.~\citep{Ali:2022guz} found that the diffraction induced by a point mass or singular isothermal sphere (SIS) lens can be identified by using a model-independent method.
However, the stochastic nature of the microlensing field poses a formidable challenge in creating a comprehensive template bank, which is capable of effectively filtering these fringes.
To address this challenge, we employ a template-free approach, known as the coherent Wave Burst (\texttt{cWB})~\citep{klimenko_sergey_2021_4419902, Klimenko:2015ypf}, to reconstruct the GW waveform. 
This method primarily involves analyzing the synchronized triggers from multiple detectors during the GW's propagation from one detector to another.
\texttt{cWB} is more suitable for finding the burst signal instead of the long duration one~\cite{Relton:2022whr}. 
Compared with BBH mergers, BNS mergers have longer durations, and NSBH merger even do not have the chirp behavior. 
Hence, in this paper, we focus on the GWs generated by BBH only. 

In this study, we introduce an interference-based approach for the identification of SLGWs. Our approach involves the detection of SLGWs, searching for pairs of SLGWs, and identifying the host galaxies. This methodology effectively addresses the inherent challenges of traditional methods. Consequently, it will enrich the utility of SLGWs in astrophysics and cosmology.

\begin{figure}[hp]
\vspace{0.2em}\centering\includegraphics[width=0.7\columnwidth]{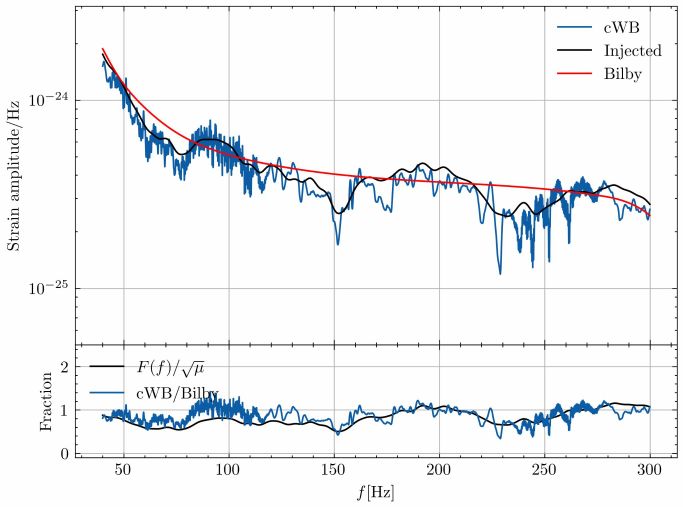}
\caption{{\bf \texttt{cWB} and \texttt{Bilby} reconstruction results}. The blue and red curves in the upper panel represent the reconstruction results of \texttt{cWB} and \texttt{Bilby}, respectively.
The black curve is the injected GW waveform.
The $x$-axis is the GW frequency, and the $y$-axis is the absolute value of the waveform.
The blue curve in the lower panel shows the ratio of \texttt{cWB} and \texttt{Bilby} results. 
The black curve is the injected ratio of the wave optics magnification factor $F(f)$ and the square root of the macro magnification $\sqrt{\mu}$. In this figure, the macro magnification is set to \(\mu = 66\), and the microlensing convergence is set to \(\kappa_{\ast} = 0.09\). The BBH parameters used in this figure are provided in Table \ref{fig:rec}.}  
\label{fig:rec}
\end{figure}

\begin{center}
{\textbf{ \Large \uppercase{results}} }
\end{center}
\noindent
{\textbf{cWB reconstruction}}
\label{sec:micro}
We illustrate our result by simulating an SLGW event generated by a BBH merger, as depicted in Figure~\ref{fig:rec}. We adopt the single-precessing-spin waveform model \texttt{IMRPhenomPv2}~\citep{Hannam:2013oca, 2020ascl.soft12021L} encoded in \texttt{PyCBC}~\citep{alex_nitz_2022_6324278} and three CE detectors located at Livingston (USA), Hanford (USA) and Pisa (Italy) to generate the simulated strain data. To illustrate the the interference effect with better visual clarity, the macro lensing magnification of the event in Figure~\ref{fig:rec} is chosen as 66 (macro lensing convergence $\kappa \simeq 0.492$, macro lensing shear $\gamma \simeq 0.492$).
Furthermore, we conservatively choose the microlensing convergence as $\kappa_* = 0.09$, which corresponds to $f_* \simeq 0.2$.
This $f_*$ value is almost the lower bound suggested by Dobler {\it et al.}~\citet{Dobler:2006wv}. The microlens mass function of the event depicted in Figure~\ref{fig:rec} differs from the one used throughout the rest of the paper, as it is chosen to be uniformly $1$ solar mass for simplicity.
The binary black hole parameters are listed in Table.~\ref{ta:BBHParm1}.

As is shown in the black curve in Figure~\ref{fig:rec}, the microlensing wave optics effect leaves a frequency-dependent imprint on the GW waveform.
Currently, while the techniques for searching this feature produced by isolated microlens have matured~\citep{Hannuksela:2019kle, LIGOScientific:2021izm, Ali:2022guz}, only a few pioneering works have studied the microlensing field scenario~\citep{Diego:2019lcd, Mishra:2021xzz, Meena:2022unp}.
The waveform template of GW intersecting with the stochastic microlensing fields could not be modeled deterministically. 
Hence, the traditional matched filtering method is no longer suitable for our goal.
Fortunately, as shown in Figure~\ref{fig:rec}, these microlensing imprints can be reconstructed using a template-free method, \texttt{cWB}.
The blue curve in the upper panel shows the reconstructed GW waveform from \texttt{cWB}.
The $x$-axis is the GW frequency, and the $y$-axis is the absolute value of  the waveform.
The blue curve is consistent with the black one, which is our injected microlensed GW signal.
The extra fast oscillations in the blue curve compared with the black is the unwanted instrumental noise. 
This result demonstrates the robustness of \texttt{cWB} for reconstructing the microlensing effect.
Furthermore, we show the best-fit waveform reconstructed from the template fitting using the template without microlensing in the smoothing red curve. 
The waveform template used in parameter estimation is \texttt{IMRPhenomPv2} encoded in \texttt{Bilby}~\citep{Ashton:2018jfp}. 
One can find that the \texttt{Bilby} result is very different from the result of \texttt{cWB}, which indicates that the fifteen parameters waveform can not reconstruct the microlensing wave optics effect at all.
The lower panel shows the ratio between $\tilde{h}_\mathrm{cWB}$ and $\tilde{h}_\mathrm{Bilby}$ as the blue curve.
By comparing with the injected value, \( F(f)/\sqrt{\mu} \), it is clear that \texttt{cWB} accurately captures the microlensing effects.

\noindent
{\textbf {Identification of SLGW signal}}
In this section, we introduce a new method for the authentication of SLGW events. 
Specifically, our approach involves the evaluation of mismatch between \texttt{cWB} and \texttt{Bilby} outcomes, serving as a means to ascertain the eligibility of a given event as an SLGW event.
One can imagine that the efficiency of this method depends on the quality of the reconstruction results and the strength of the microlensing imprints.
To demonstrate the reliability of the above method, we need to know the extent to which unlensed events can mimic the result of lensed events.
We randomly select $200$ unlensed GWs to construct the false positive sets (see \textbf{METHOD} for details).

\begin{figure}[hp]
\vspace{0.2em}\centering\includegraphics[width=0.7\columnwidth]{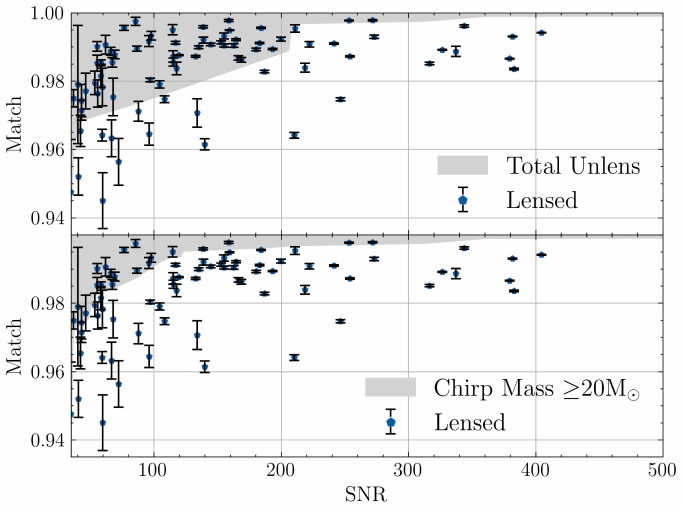}
\caption{\textbf{Identification of SLGW events}. 
This figure shows the match between \texttt{cWB}'s maximum likelihood waveform and \texttt{Bilby}'s posterior results, as a function of SNR.
The shaded areas in grey delineate the envelope of the lower matching value between the maximum likelihood waveform of \texttt{cWB} and the posterior results from \texttt{Bilby} across all unlensed events (false positive samples).
The black error bars ($90\%$ confidence interval) with blue pentagrams (mean value) represent the match results of SLGWs.
The upper and lower panels show the results without and with detector frame chirp mass $\mathcal{M}_z\ge 20M_{\odot}$ cut, respectively.
Our simulation is conducted by assuming three CE detectors over three years.}
\label{fig:match}
\end{figure}

The grey shaded areas in Figure~\ref{fig:match} represent the match result of \texttt{cWB} and \texttt{Bilby} for these false positive samples.
The $x$-axis stands for the matched-filter SNR.
It is worth mentioning that when calculating the matching for each event, we randomly select $100$ groups of parameter values from the posterior distribution of the \texttt{Bilby} results and match them with the best-fit result of \texttt{cWB}.
The envelope of the shaded area is the lower matching bound of all false positive events.
The upper and lower panels stand for results without and with detector frame chirp mass $\mathcal{M}_z=20~\mathrm{M}_\odot$ cut, respectively.
One can find that the match value is proportional to SNR.
This result is expected because, at high SNR, both \texttt{cWB} and \texttt{Bilby} can faithfully reconstruct the actual GW waveform with tiny uncertainty. This is consistent with the result from another \texttt{cWB} reconstruction work~\cite{bini2024searchhyperbolicencounterscompact}.
Comparing the two panels demonstrates that $\mathcal{M}_z > 20~\mathrm{M}_\odot$ truncation can significantly improve the matching result for events with SNR$\in(40,200)$. 
We note that setting a cutoff of $\mathcal{M}_z = 20~\mathrm{M}_\odot$ is cost-effective. 
It only loses approximately $17 \%$ SLGWs, but can significantly reduce the FAP. 

Based on our simulation, we expect to detect $510$ SLGWs with an SNR greater than $12$ over a 3-year period, originating from $256$ strong lensing systems. Of these, we estimate that $85$ signals exhibit strong microlensing signatures. 
These events are plotted as the blue pentagrams (mean value) with black error bars ($90\%$ confidence interval)  in Figure~\ref{fig:match}. 
They are identified by comparing the match between the theoretical input signals with and without microlensing effects. More specifically, we select events where the theoretical match is below $99.5\%$ for signals with SNR$<150$, and below $99.99\%$ for signals with SNR$>150$. Since these thresholds closely align with the boundary of the shaded region, we do not expect the remaining $425$ events, which exhibit only weak microlensing effects, to be distinguishable.
Subsequently, we conduct parameter estimation for each of the $85$ events using \texttt{Billy} and \texttt{cWB}. Among these, $58$ are classified as microlensing identifiable events, where the upper error bars do not overlap with the lower boundary of the shaded region. Consequently, we conclude that $27$ events are missed due to estimation uncertainties. In summary, our method has the capacity to identify more than $10\%$ ($58$ out of $510$) of SLGWs.

\noindent
{\textbf{Strong lensing pairing}}
In the preceding section, we successfully authenticated $58$ single-image SLGWs every three years. 
Through an analysis of the sky localization overlapping between these $58$ single-image and the remaining GWs, we are able to select the multiple-image systems associated with these single-images.

\begin{figure}[hp]
\vspace{0.2em}\centering\includegraphics[width=1\columnwidth]{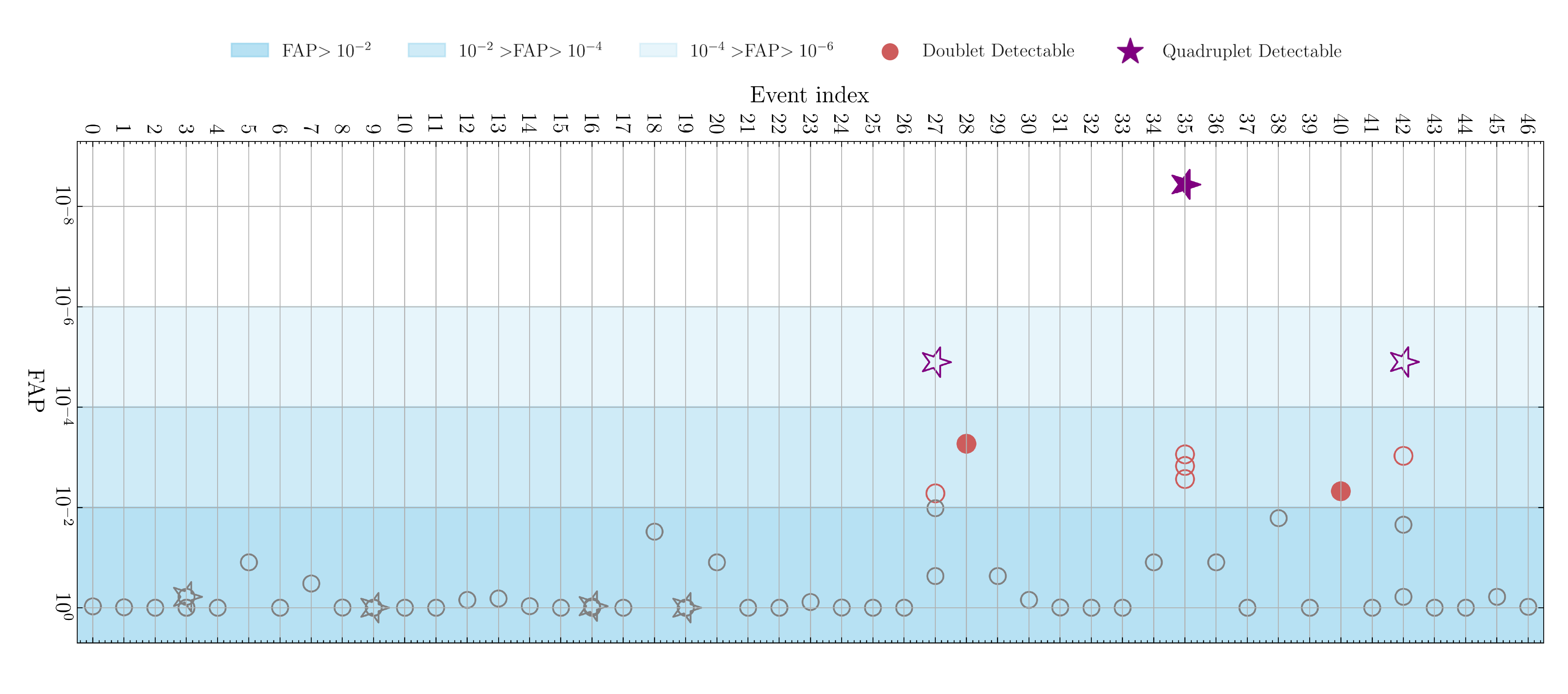}
\caption{\textbf{Identification of SLGW pairs}. This figure displays the results of finding SLGW pairs. 
The $y$-axis represents the FAP, defined in Eq. (\ref{eq:trial_factor}) for doublet and Eq. (\ref{eq:FAP_system}) for quadruplet. 
The $x$-axis corresponds to the event index. 
The divide between different color regions corresponds to $\text{FAP}=10^{-2}, 10^{-4},$ and $10^{-6}$, respectively.
The circles and stars indicate the FAP of doublets and quadruplets, respectively.  Grey represents $\text{FAP}>0.01$, red represents $10^{-4}<\text{FAP}<0.01$, and purple represents $\text{FAP}<10^{-4}$. 
We use a successful identification threshold of  FAP$<10^{-2}$ for doublets and  FAP$<10^{-6}$ for quadruplets. Applying these criteria, we identified $2$ double-image systems and $1$ quadruple-image system, which are represented by solid circles and a solid star, respectively.
}
\label{fig:identified_pair}
\end{figure}

Figure~\ref{fig:identified_pair} presents the results of our multiple-image identification process. 
We always pair the GWs with the first detected signal among multiple-images. Therefore, a double-image corresponds to 1 pair, and a quadruple-image corresponds to 3 pairs. 
The $y$-axis represents the FAP including the trial factor derived from $10^5$ false positives according to Eq. (\ref{eq:trial_factor}). 
The $x$-axis corresponds to the event index. 
It is worth noting that each of the previously identified $58$ images belong to either a new lens system or an old system shared with other identified images. 
In general, these $58$ images are included in $47$ strong lensing systems. 
The divide between different color regions corresponds to $\text{FAP}=10^{-2}, 10^{-4},$ and $10^{-6}$, respectively.
The circles in the figure indicate the FAP of a doublet and stars denote the FAP of a quadruplet, which is defined in Eq.~(\ref{eq:FAP_system}).  
Grey represents $\text{FAP}>0.01$, red represents $10^{-4}<\text{FAP}<0.01$, and purple represents $\text{FAP}<10^{-4}$. 
In this paper, we adopt a threshold of FAP$<10^{-2}$ for doublet and FAP$<10^{-6}$ for quadruplet. 
With this choice, $2$ double-image and $1$ quadruple-image systems were identified, which are marked as solid circles and solid star. 
Particularly, for the quadruple event ID-35, its FAP is \(<10^{-8}\), with each image pair having a FAP of \(<10^{-2}\).

\noindent
{\textbf{Host galaxy identification}}
In the context of quadruple-image systems, the identification of host galaxies can be accomplished through a comparison between the time-delays of SLGW and galaxy-galaxy strong lensing (GGSL) events, as detailed in Hannuksela {\it et al.}~\citep{Hannuksela:2020xor} 
For double-image system, it is difficult to pinpoint the host galaxy. Hence, we do not analyse double-image system at this step.
For quadruple-image system, the BBH must reside in the area of source galaxy, which is overlapped with the caustics.  
Statistically, the area which can generate the consistent time-delay with those from GW, shall be proportional to the probability of the source galaxy being the host.
Hereafter, we will call this area as ``time-delay area''.  
Considering the BBH population model, we need to further weight this area according to its star formation rate (SFR).

\begin{figure}[hp]
\vspace{0.2em}\centering\includegraphics[width=0.5\columnwidth]{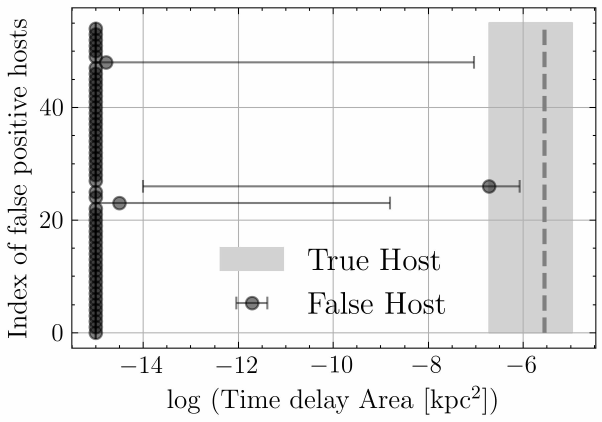}
\caption{\textbf{Identification of SLGW Host galaxy}. 
This figure displays the results of host galaxy identification. 
The $x$-axis represents the logarithmic value of the ``time-delay area'', where the time-delay ratio is within $1\text{\textperthousand}$ agreement with those from SLGW.
The $y$-axis denotes the event index of the false GGSL. 
The grey vertical dashed line and the dark grey circles represent the average areas for true host galaxies and false host galaxies, respectively.
Both the light grey shaded region and the error bars of the dark grey circles denote the $1\sigma$ confidence intervals.
}
\label{fig:Host_galaxy}
\end{figure}

Figure~\ref{fig:Host_galaxy} showcases the host galaxy identification result of quadruple event ID-$35$ in Figure~\ref{fig:identified_pair}, acquired using one of the flagship stage-IV galaxy survey, namely China Space Station Telescope (CSST)~\citep{2011SSPMA..41.1441Z} and James Webb Space Telescope (JWST)~\citep{Gardner:2006ky}. 
CSST is used to select the GGSL candidates thanks to its wide field of view. And JWST is used for a dedicated follow-up. 
Among the three purple quadruplets shown in Figure \ref{fig:identified_pair}, the host galaxy of the ID-35 event stands out as the brightest (smallest source redshift, $z_s=1.6$), as demonstrated in Extended Data {\bf Fig. 1}, 
and most accurately localized ($1.3$ square degrees) one.
It is worth noting that the $1.3$ square degrees is the sky localization envelope region, rather than the overlapping region, of the multiple GW counterparts.  
In Extended Data {\bf Fig. 2}, 
we show the sky localization result for three quadruplets, highlighted in purple in Figure~\ref{fig:identified_pair}. 
Each panel has four counters representing quadruple counterparts.
The injected sky location (dashed curve) is safely within the envelope of the sky localization.

The $x$-axis of Figure~\ref{fig:Host_galaxy} represents the logarithmic value of the ``time-delay area''. 
The grey vertical dashed line represents the average area for true host galaxies, while the light grey shaded region indicates the uncertainties, which account for variations in both the properties of the host galaxies and the positions of the BBHs within them. We randomly select $40$ different host galaxies with different magnitudes, spectral energy distributions, and light S\'{e}rsic profiles based on JWST mock catalog at the same redshift. For each host galaxy, we randomly sample $100$ spatial positions to compute the time-delays, accounting for the uncertainty in the exact position of the BBH.
The dark grey circles with errors represent the false hosts. 
Both the shaded region and error bars mark $1\sigma$ confidence intervals. 
In order to obtain a reliable statistics, we included all the GGSL systems (number is 54) which pass the CSST criteria, within $20$ square degrees instead of $1.3$ square degrees. 
As been demonstrated previously, the ``true host galaxy shall have the largest area". 
According to our simulation, for event ID-35, the average confidence, as defined in Eq.(\ref{eq:host_sigma}), under the hypothesis that the ``true host galaxy has the largest area" is approximately \( 7.75 \, \sigma \). This means that our method allows us to confidently identify the true host galaxy.
So far, We have presented all the essential components of our method. The efficiency of SLGW identification through each steps are summarized in Table~\ref{ta:all_ide}.

\begin{center}
{\textbf{ \Large \uppercase{Discussion}} }
\end{center}
Aside from false positive events caused by random noise, another important concern is the potential degeneracy with spin precession.
To explore this, we simulated another $200$ unlensed signals with precession, where the probability density function of the spin orientation is uniformly distributed in spherical coordinates. 
As shown in Extended Data {\bf Fig. 3}, the precession effect only affects identification efficiency at low SNR. 
Specifically, for SNR $<50$, the match decreases from $0.97$ to $0.95$, while for SNR $>50$, precession has no noticeable effect. After accounting for spin precession, we lose only $2$ out of $58$ microlensing identifiable events with SNRs below $50$. 

The phenomenological differences between the spin precession effect and the microlensing diffraction imprint are clear: the spin precession effect evolves gradually and smoothly over time, while the microlensing field exhibits more erratic, random fluctuations, particularly at higher frequencies.
To substantiate this argument, we present in Extended Data {\bf Fig. 4} the waveform of one of the identifiable events from Figure~\ref{fig:match}. 
Its macro magnification is $2.2$, the SNR is $179$, and the match value is $0.987$. 
The orange curve corresponds to the injected GW waveform,  which includes microlensing but excludes precession. 
The grey and blue curves represent the maximum likelihood reconstruction results from \texttt{cWB} and \texttt{Bilby}, respectively. 
In \texttt{Bilby}, we choose a precession template, namely \texttt{IMRPhenomPv2}. 
The first panel displays the waveform in the frequency domain, with the $x$-axis representing the GW frequency and the $y$-axis representing the amplitude. 
It is evident that the grey curve provides a better fit to the orange curve than the blue one. 

The differences between microlensing and precession become more evident when examined in the time domain.
The second panel provides a zoomed-in view of the time domain waveform from merger phase. 
One can see that the precession waveform fails to capture certain high-frequency modulations produced by microlensing.
The third and fourth panels display the full zoomed-out waveform, starting from $10$ Hz.
Precession clearly induces low-frequency modulation during the inspiral phase, noticeable after the red vertical line in the third panel. In contrast, microlensing shows no such effect, as evident in the fourth panel.
Therefore, precession alone is unable to replicate the interference imprint caused by microlensing.
However, this does not mean there is no leakage from microlensing into precession, especially in events with strong microlensing signatures. 
In Extended Data {\bf Fig. 5}, we present the posterior distribution of the effective precession spin parameter~\cite{Schmidt_2015}, $\chi_p$, for the same events shown in Extended Data {\bf Fig. 4}. 
It is obvious that the distribution deviates from zero, indicating the presence of the precession leakage. 

One might question whether the ID-35 quadruple event is indeed a very special occurrence, to the extent that its discovery was purely accidental. 
To address this issue, we conducted simulations of SLGWs over $30$ years. 
The results are shown in Extended Data {\bf Fig. 6}.
We found that $3$ CE detectors can identify $91$ out of $516$ signals in quadruple-image systems. 
These $91$ identifiable signals and the total $516$ signals are included in $38$ and $129$ quadruple-image systems, respectively. 
Extended Data {\bf Fig. 7}
illustrates the redshift and sky localization of these $38$ quadruple-image systems. 
We found that there are $18$ quadruple-image systems below $z_s < 2.1$, with sky localization areas under $5$ square degrees. 
For them, CSST has more than $60\%$ probability to observe its host galaxy. 
Therefore, the ID-$35$ event is not a special event by coincidence, and our proposed method is robust for identifying SLGWs and associated host galaxies.

Furthermore, it is important to note that the identification of GGSL associated with SLGW could be even more promising. 
In this analysis, we choose the space-borne telescopes CSST and JWST for the strong lensing image observation. 
While space-borne telescopes have more accurate angular resolution, their limiting magnitude is lower compared to large ground-based telescopes.
This limitation fails to find the fainter events, such as ID-27 and ID-42. 
To address this challenge, we propose to use large ground-based survey telescopes, such as the Rubin Observatory \cite{LSSTScience:2009jmu,Smith:2019qsv}, to identify GGSL systems.
Subsequently, employing smaller field of view telescopes equipped with adaptive optical systems, like the Thirty Meter Telescope \cite{TMTInternationalScienceDevelopmentTeamsTMTScienceAdvisoryCommittee:2015pvw}, to conduct precise follow-up observations. 
The combined use of these instruments can further enhance our ability to identify the host galaxies.
In summary, we proposed a promising identification method for SLGW and associated host galaxy, triggered by the microlensing wave optics. We have validated that it is robust against all the uncertainties we were concerned about.

\begin{center}
{\textbf{ \Large \uppercase{method}}}
\end{center}

\noindent
{\textbf{SLGW mock data simulation}}
To validate the method, we follow Refs.~\citet{Haris:2018vmn,Xu:2021bfn} to generate a mock data set consisting of both lensed and unlensed data using the Monte Carlo method.
The primary simulation process is as follows. 
\begin{itemize}
  \item [1.] 
  We sample the BBH redshift from a theoretical BBH merger rate model in which the merger rate is proportional to the SFR with a delay time $\Delta t = 50 \mathrm{Myr}$ between the star and BBH formation. The details can be found in Appendix B of Xu {\it et al.}~\citet{Xu:2021bfn}.
  \item [2.]
  For the events picked above, we randomly assign BBH masses ($m_1$, $m_2$), inclination angle ($\iota$), polarization angle ($\psi$), right ascension angle ($\alpha$), declination ($\delta$), merger time ($t_c$), and spins ($a_1$, $a_2$) from the following distributions. 
      \begin{itemize}
      \item [a)]
      $(m_1, m_2)\sim \bf\mathrm{power\ law + peak}$~\citep{LIGOScientific:2018jsj}.
      \item [b)]
      $p(\iota)\propto \sin(\iota)$, $\iota \in  [0, \pi]$.
      \item [c)]
      $p(\psi)\propto \mathrm{U}(0,\pi)$.
      \item [d)]
      $p(\alpha)\propto \mathrm{U}(0,2\pi)$.
      \item [e)]
      $p(\delta)\propto \cos(\delta)$, $\delta \in [-\pi/2, \pi/2]$.
      \item [f)]
      $p(t_c)\propto \mathrm{U}(t_\mathrm{min}, t_\mathrm{max})$, where $t_\mathrm{min}$ and $t_\mathrm{max}$ are the minimum and maximum merger times used in the simulation. Here, we set $t_\mathrm{max}-t_\mathrm{min}=3\mathrm{yr}\times 80\%$(duty cycle).
      \item [g)]
      $p(a_1)\propto \mathrm{U}(0, 0.99)$.
      \item [h)]
      $p(a_2)\propto \mathrm{U}(0, 0.99)$.
      \end{itemize}
  \item [3.]
  Calculate the multiple-image optical depth $\tau(z_s)$ for each BBH redshift $z_s$ using the SIS optical depth as shown in Haris {\it et al.}~\citep{Haris:2018vmn}. 
  Then, generate a random number uniformly distributed between $0$ and $1$ for each BBH event. 
  Compare the calculated optical depth $\tau(z_s)$ with the generated random number for each event. 
  If the optical depth $\tau(z_s)$ is greater than the random number, classify it as an SLGW event; otherwise, exclude it from the selection. 
   \item [4.]
  For the selected SLGW samples, we assume a SIE lens model~\citep{1994A&A...284..285K} and use \texttt{Lenstronomy}~\citep{2018PDU....22..189B, 2021JOSS....6.3283B} to solve the lens equation.
  The velocity dispersion $\sigma_v$ and axis ratio $q$ of SIE are generated from the SDSS galaxy population distribution~\citep{2015ApJ...811...20C}. 
  Note that Ref.~\citet{2015ApJ...811...20C} has a typo in axis ratio parameter, we use the corrected form in Ref.~\citet{Wierda:2021upe}. 
  The sample details for these parameters, lens redshift, and source-plane location can be found in Appendix A of Haris {\it et al.}~\citet{Haris:2018vmn}.
\end{itemize}

After accounting for the detector's selection effect in the provided samples, three CE detectors, located at Livingston (USA), Hanford (USA) and Pisa (Italy), can potentially observe approximately $3.3\times10^{5}$ BBHs and $510$ SLGWs ($256$ strong lensing systems) in $3$ years with $80\%$ duty cycle. 
This result aligns with the findings of Xu {\it et al.}~\citet{Xu:2021bfn}.
It's important to note that in this simulation, we assume that an event will be considered as a detection if it possesses a network matched filter signal-to-noise ratio (SNR) $\geq 12$.
Additionally, it's worth highlighting that, despite using three CE detectors in this simulation, we calculate the SNR starting from a frequency of $20$Hz, not from $1$Hz, attributed to computational constraints.
Therefore, the result is conservative.

Now, our focus shifts to the simulation of microlensing field, following the recipe listed in Refs.~\cite{Chen:2021ftm, Zheng:2022vfq,Shan:2022xfx}. 
In this study, we utilize the Salpeter initial mass function~\citep{1955ApJ...121..161S} and an elliptical S\'{e}rsic profile~\citep{Vernardos_2018} to describe the stellar mass function and density associated with each SLGW.
Specifically, we set the stellar mass range to be within $[0.1, 1.5]$ solar masses, which aligns with the value employed by Diego {\it et al.}~\citep{Diego:2021mhf}.
In addition to the stellar mass component, we also consider the presence of remnant objects in the microlensing field. 
For this purpose, we adopt the initial-final relation from Spera {\it et al.}~\citet{2015MNRAS.451.4086S}.
The remnant mass density has been set at $10\%$ of the stellar mass density~\citep{Meena:2022unp}.

To provide the frequency dependent magnification, we use the algorithm introduced in Shan {\it et al.}~\citet{Shan:2022xfx} to evaluate the Fresnel-Kirchhoff diffraction integral~\citep{1992grlebookS}
\begin{equation}
\label{eq:DiffInter}
F(\omega, \boldsymbol{y})=\frac{2 G \mathrm{M}_{L}\left(1+z_{L}\right) \omega}{\pi c^{3} i} \int_{-\infty}^{\infty} d^{2} x \exp \left[i \omega t(\boldsymbol{x}, \boldsymbol{y})\right]\;,
\end{equation}
where $F(\omega, \boldsymbol{y})$ is the wave optics magnification factor, $\omega$ and $\boldsymbol{y}$ are the circular frequency of the GW and its position in the source plane in the unit of the Einstein radius. 
$ \mathrm{M}_{L}$ and $z_L$ are the lens mass and redshift, $\boldsymbol{x}$ is the lens plane coordinate, and $t(\boldsymbol{x}, \boldsymbol{y})$ is the time-delay function defined as 
\begin{equation}
\begin{split}
\label{eq:TimeDelay}
t(\boldsymbol{x},\boldsymbol{x}^{i},\boldsymbol{y}=0)&=\underbrace{\frac{k}{2}\left((1-\kappa+\gamma) x_{1}^{2}+(1-\kappa-\gamma) x_{2}^{2}\right)}_{t_\text{s}(\kappa,\gamma,\boldsymbol{x})} -  \underbrace{\left[\frac{k}{2}\sum_{i}^{N} \ln \left(\boldsymbol{x}^{i}-\boldsymbol{x}\right)^{2} + k\phi_{-}(\boldsymbol{x})\right]}_{t_\text{m}(\boldsymbol{x},\boldsymbol{x}^{i})}
\end{split}
\end{equation}
where $k=4 G \text{M}_\text{micro}(1+z_L)/c^3$ and $\boldsymbol{x^{i}}$ is coordinate of the $i$th microlens. The parameter \(\text{M}_\text{micro}\) represents the average microlensing mass. It is set to $1$ solar mass in Figure~\ref{fig:rec} and $0.35$ solar mass in the rest of the paper.

Here, we set the macro image point as the coordinate origin ($y = 0$).
$\phi_{-}(\boldsymbol{x})$ is the contribution from a negative mass sheet which is used to cancel out the mass contribution from microlenses and keep the total convergence $\kappa$ unchanged~\citep{Wambsganss1990, Chen:2021ftm, Zheng:2022vfq}.
$t_\text{s}(\kappa, \gamma, \boldsymbol{x})$ represents the macro lens time-delay and $t_\text{m}(\boldsymbol{x},\boldsymbol{x}^{i})$ indicates the microlens time-delay. Up to this step, we have successfully generated all the essential components for the GW mock data, encompassing both unlensed GWs and SLGWs with microlensing effects.

\noindent
{\textbf{SLGW finder and pairing}}
The mismatch between \texttt{cWB} and \texttt{Bilby} serves as a mean to find SLGWs. Here, we define the match equation as
\begin{equation}
\label{eq:match}
\mathrm{match} = \frac{\left\langle \tilde{h}_\mathrm{cWB} \mid \tilde{h}_{\mathrm{Bilby}}\right\rangle}{\sqrt{\left\langle \tilde{h}_\mathrm{cWB} \mid \tilde{h}_\mathrm{cWB}\right\rangle\left\langle \tilde{h}_{\mathrm{Bilby}} \mid \tilde{h}_{\mathrm{Bilby}}\right\rangle}} \;,
\end{equation}
where $\tilde{h}_\mathrm{cWB}$ and $\tilde{h}_\mathrm{Bilby}$ are the reconstructed waveforms in the frequency domain.
$\left\langle . \mid . \right\rangle$ stands for the noise-weighted inner product and is defined as
\begin{equation}
\left\langle \tilde{h}_{1} \mid \tilde{h}_{2}\right\rangle=4 \operatorname{Re} \int_{f_{\text {low }}}^{f_{\text {high }}} \mathrm{d} f \frac{|\tilde{h}_{1}(f)|\times |\tilde{h}_{2}(f)|}{S_{\mathrm{n}}(f)} \;,
\end{equation}
where $|.|$ refers to the absolute value, and $S_\mathrm{n}(f)$ is the single-side power spectral density of the detector noise.
It is evident that Eq.~(\ref{eq:match}) is $\leq 1$, and the equality holds if and only if $\tilde{h}_\mathrm{cWB} = \tilde{h}_\mathrm{Bilby}$.

We search for SLGW multiple-image pairs based on the parameter overlapping degree between two GW events. 
To do this, we utilize the ``overlapping'' method introduced in Haris {\it et al.}~\citep{Haris:2018vmn}. 
\begin{equation}
\mathcal{B}_{\mathrm{U}}^{\mathrm{L}}:=\int d \boldsymbol{\theta} \frac{P\left(\boldsymbol{\theta} \mid d_{1}\right) P\left(\boldsymbol{\theta} \mid d_{2}\right)}{P(\boldsymbol{\theta})} \\,
\end{equation}
where $\theta$ represents the GW parameter, $d_1$ and $d_2$ denote the strain data for event $1$ and event $2$, respectively.
$P(\boldsymbol{\theta})$ corresponds to the prior distribution, and $P\left(\boldsymbol{\theta} \mid d_{1} (d_{2})\right)$ represents the posterior distribution.
In this calculation, we only consider two parameters, RA (right ascension) and DEC (declination). 
This choice is motivated by the fact that the presence of the microlensing effect does not introduce significant bias on these two parameters.

To demonstrate the identification accuracy of the pairing method, it is crucial to assess the FAP.
First, we define the FAP per pair as
\begin{equation}
\label{eq:FAP_pair}
\mathrm{FAP}_{\rm per~pair}=\frac{N_\mathrm{UU+UL}(\mathcal{B}>\mathcal{B}_\mathrm{L})}{N_\mathrm{UU+UL}(\mathrm{total})} \\.
\end{equation}
The numerator is the number of false positives. The Bayes factor of these false positives $\mathcal{B}$ are higher than the Bayes factor of SLGW image pair $\mathcal{B}_\mathrm{L}$. The denominator is the total number of randomly matched unlensed pairs and unlens-lens pairs. 
For doublet, the FAP after including the trial factor is defined as~\citep{Caliskan:2022wbh}
\begin{equation}
\label{eq:trial_factor}
    \mathrm{FAP} = 1 - (1 - \mathrm{FAP}_\mathrm{per\ pair})^{N_{\rm per~year}}\;,\;\;(\rm{doublet})\;.
\end{equation}
It depends exponentially on the number of pairs. In our method, \(N_{\rm per~year}\) represents the number of detectable GWs per year. For third-generation GW detectors, we select \(N_{\rm per~year}=10^5\). In contrast, without utilizing microlensing information, the exponential term becomes \(N_{\rm UU+UL} \simeq N_{\rm per~year}^2\). Therefore, one can conclude that our method significantly reduces the FAP. 

To estimate the FAP of a quadruplet, we simply take the product of the FAPs of three individual doublets
\begin{equation}
\label{eq:FAP_system}
    \mathrm{FAP} = \mathrm{FAP}_1 \times \mathrm{FAP}_2 \times \mathrm{FAP}_3\;,\;\;(\rm{quadruplet})\;.
\end{equation}
This estimator offers a computationally simple and mathematically conservative way to calculate the FAP for a quadruplet. It is based on the overlap between the individual doublets within the quadruplet, rather than requiring all four images to overlap simultaneously. This condition is less stringent, making our result more conservative.

\noindent
{\textbf{GGSL simulation and host galaxy identification}}
In this section, we introduce our host galaxy identification method for SLGWs. 
We first generate a mock dataset for GGSL by utilizing a JWST mock catalog known as JAGUAR~\citep{2018ApJS..236...33W}.
For the false GGSL systems, we employ the optical depth method, which is identical to the one used for generating SLGWs, to simulate GGSL events across a $20$ square degrees region. 
We find that there are roughly $3300$ GGSL systems with Einstein radius $\theta_E>0.2''$ in $1$ square degree. 
This number is consistent with the simulation result of the CSST strong lensing group. 
Subsequently, we randomly select lens galaxy magnitudes and light S\'{e}rsic radius. Note that there is a typographical error in Goldstein {\it et al.}\citep{Goldstein:2018bue}, so we utilize the corrected formula provided in Wempe {\it et al.}~\citep{Wempe:2022zlk}. using the fundamental plane~\citep{Goldstein:2018bue}.
For the host galaxy, we collect the galaxy properties, such as spectral energy distribution and light S\'{e}rsic profile, via a thin shell $[z_s - \Delta z_s, z_s + \Delta z_s]$, where $z_s$ is the real host galaxy redshift and the shell width is chosen as $\Delta z_s=0.01$. 
The true host galaxy property parameter is assigned according to the above samples. 
We then rank the host probability based on the SFR of each samples over the past $50~\mathrm{Myr}$.

In order to find the host galaxies, we propose a targeted observation strategy. 
First, we conduct an ordinary survey ($600s$ exposure time) utilizing the CSST, which has a field of view around $1.1$ square degrees.
The primary objective of this step is to systematically scan the sky localization envelope of multiple-image SLGWs and subsequently select the GGSL systems which are observable.
Here, we employ two criteria to assess the observability of GGSL systems: $M_\mathrm{AB} < 26$, and $\theta_{\mathrm{E}}^{2}>r_{s}^{2}+(s / 2)^{2}$, where $\theta_\mathrm{E}$ represents the Einstein radius, $s$ denotes the seeing (for CSST $s=0.135^{''}$), and $r_s$ stands for the unlensed source size.
The second criteria denotes the requirement of being able to distinguish multiple images in the GGSL system.

Subsequently, we propose to use JWST, which has a larger aperture than CSST, for dedicated follow-up observations for each of the targeted GGSLs. We propose a $1000s$ exposure for each of the targets. According to the JWST Exposure Time Calculator, \url{https://jwst.etc.stsci.edu}, an exposure time of $1000s$ yields an SNR $>33$ for a point source with magnitude $<26$~\cite{Pickering_2016} in F$200$W band. The choice ensures the quality of lens image reconstruction.
This strategy is cost-effective since CSST observation will only select around $3$ quadurple-image GGSLs per square degree. 
Hence, the subsequent JWST observations time is about $1$ hour in total for $3$ candidates. 

In Extended Data {\bf Fig. 1}, we depict the probability distribution of host galaxy apparent magnitudes for the three quadruplets. 
The host galaxy number density is weighted by the SFR according to the BBH population model. 
In this figure, red histogram is the apparent magnitude distribution for CSST $r$ band, and the blue histogram is the one for JWST F$200$W band. 
The difference between the red and blue only results from the filters and SED, nothing to do with the telescope aperture and exposure time. 
It is worth noting that our current analysis assumes only single photometry band, and the multi-band analysis will definitely improve the current results. 
The grey shaded region indicates events that cannot be observed by CSST due to its limited magnitude (assuming CSST limiting magnitude of $\mathrm{M}_{AB}=26$).
From this figure, it is clear that for event ID-35, there is a remarkably high probability (approximately $80\%$) of being able to observe its host galaxy by CSST. 

To identify host galaxies, we ask for the consistency of time-delays between GGSL and SLGW measurements. 
For quadruple-image systems, the estimator consists of two independent components: $\Delta t_{1,2} / \Delta t_{1,3}$ and $\Delta t_{1,2} / \Delta t_{1,4}$.
Here, $\Delta t_{1,2}$ represents the time-delay between image $1$ and $2$ (with $\Delta t_{1,3}$ and $\Delta t_{1,4}$ having similar meanings).
In detail, the estimator is defined as
\begin{equation}
\label{eq:discriminator}
A_\mathrm{con} = \sqrt{A_\mathrm{GGSL}(\frac{\Delta t_{1,2}}{\Delta t_{1,3}})|_\mathrm{SLGW} \times A_\mathrm{GGSL}(\frac{\Delta t_{1,2}}{\Delta t_{1,4}})|_\mathrm{SLGW} \times W^2_\mathrm{SFR}}.
\end{equation}
$A_\mathrm{GGSL}(x)|_\mathrm{SLGW}$ represents the area (in unit of kpc$^2$) in the source galaxy, in which each of the pixels can generate the time-delay ratio agreeing with those from SLGW within $1\text{\textperthousand}$ precision. 
We also tested the convergence of the result by using the precision of $10^{-4}$. 
Below $10^{-4}$, we can not resolve single pixel in our simulated lensing image anymore.
Furthermore, we require the absolute time-delay between image $1$ and $2$ to be consistent with those from GWs in the range of $(\frac{67.74}{60}\Delta t^{\rm GW}_{1,2},\frac{67.74}{80}\Delta t^{\rm GW}_{1,2}$), where $\Delta t^{\rm GW}_{1,2}$ is the GW's time-delay between image $1$ and $2$. 
The numerical factor preceding \(\Delta t^{\rm GW}_{1,2}\) accounts for the uncertainty in the Hubble parameter, which lies between $60$ and $80$ km/s/Mpc.
Our fiducial Hubble paremeter value is $67.74$ km/s/Mpc.
It is evident that the lager this area is, the greater the probability of this galaxy to be the true host. 

To incorporate with the BBH population model, we weight the pixels by their SFR ($W_\mathrm{SFR}$). 
Supplementary Figure 9
illustrates the relative positions of the host galaxy and caustic for one of the SLGW systems. 
The red curve represents the caustic of a lens galaxy, while the elliptical region indicates the half-light radius of a source galaxy, with the color (from blue to yellow) representing the source light flux (from weak to strong) distribution. 
The shaded region represents the quadruple-image region in the source galaxy.

The confidence of the hypothesis of ``true host galaxy has the largest area'' against the simulation data is defined as
\begin{equation}
\label{eq:host_sigma}
\text{Confidence} = \frac{1}{N} \Sigma_i^N \frac{\bar{A}_\mathrm{con,\ host} - \bar{A}_\mathrm{con, \ i}}{\sqrt{\Big\langle \sigma^2(A_\mathrm{con,\ host}) + \sigma^2(A_\mathrm{con, \ i})\Big\rangle}} \\,
\end{equation}
where $A_\mathrm{con,\ host}$ and $A_\mathrm{con,\ i}$ are the ``time-delay area'' for host and false hosts, defined in Eq.~(\ref{eq:discriminator}). 
The angle bracket denotes the average over 40 realizations.
The term $\frac{1}{N} \Sigma_i^N$ represents the average over all false hosts, where $i$ denotes the $i$th false host and $N$ is the total number of false hosts. This formula represents the theoretical average confidence level of the hypothesis of ``true host galaxy has the largest area''.

Up to this point, we have introduced all the simulation procedures and methods. 
To provide a clearer representation, we illustrate the main steps of our methodology in 
Supplementary Figure 10.

\vspace{5mm}
\noindent
{\textbf{Data availability}} 
The simulated microlensing data is publicly available at BNU cloud \url{https://pan.bnu.edu.cn/l/X1QPKG}. Other datasets are available at GitHub repository \url{https://github.com/xkshan97/Micro_Interference4SLGW_identification.git}.

\vspace{2mm}
\noindent
{\textbf{Code availability}} The code that support the findings of this study are available at \url{https://github.com/xkshan97/Micro_Interference4SLGW_identification.git}.

\vspace{2mm}

\noindent
{\textbf{Acknowledgements}}
This work is supported in part by the National Key R\&D Program of China No.2021YFC2203001, No.2020YFC2201502 and No.2021YFA0718304, and supported  in part by the National Natural Science Foundation of China Grants No.11821505, No.11991052 and No.12235019. 

\vspace{2mm}

\noindent
{\textbf{Author contributions}}
All authors provided ideas throughout the project and comments on the manuscript.
XS contributed in calculating and writing the draft.
BH contributed in proposing the idea and writing the draft.
XC contributed in generating the microlensing fields. 
RGC contributed in proposing the idea and writing the draft.

\vspace{2mm}

\noindent
{\textbf{Competing interests}}
The authors have no competing interests.

\begin{center}
{\textbf{ \Large \uppercase{tables}} }
\end{center}
\noindent

\begin{table*}[hp]
  \centering
  \caption{{\bf \label{ta:BBHParm1} Binary black hole parameters for Figure~\ref{fig:rec}}.
  $q$ is the mass ratio, $\mathcal{M}_\mathrm{obs}$ is the chirp mass, $z_s$ is the redshift of the source, $a_1$ is the spin magnitude of the primary black hole, $a_2$ is the spin magnitude of the secondary black hole, $\theta_{jn}$ is the inclination, ra is the right ascension, dec is the declination, $\Psi$ is the polarization angle.} 
   \begin{tabular}{ccccccccc} 
    \hline
    $q$ & $\mathcal{M}_\mathrm{obs}$ & $z_s$ & $a_1$ & $a_2$ & $\theta_{jn}$ & ra & dec & $\Psi$ \\ 
    \hline
    $0.4$ & $28.8$ & $1$ & $0.98$ & $0.46$ & $2.22$ & $5.52$ & $0.57$ & $0.27$ \\
    
    \hline
  \end{tabular}
\end{table*}

\begin{table*}[hp]
\centering
\caption{ 
{\bf \label{ta:all_ide} Detection efficiency at each steps.}
This table summarizes the number (efficiency) of SLGWs identified through each steps.
``Simulation input'' refers to the number of SLGW detected by three CE detectors over three years. 
``Single'' refers to events where only one image has an SNR greater than $12$, while ``double'', ``triple'', and ``quadruple'' correspond to systems with two, three, and four images, respectively, each with an SNR exceeding $12$.
``Total systems'' refers to the total number of strong lensing systems, and ``Total images'' refers to the total number of strong lensing images within these systems.
In summary, at Step-1 (SLGW identification), $58$ strong lensing images were identified, including $40$ double-image systems and $7$ quadruple-image systems.
At Step-2 (SLGW pairing), $2$ double-image systems (FAP$<10^{-2}$) and $1$ quadruple-image systems (FAP$<10^{-6}$) were identified.
Host galaxy identification is the Step-3.} 
\begin{tabular}{ccccccc}
\hline
                         & Single & Double         & Triple & Quadruple       & Total systems  & Total images    \\ \hline
Simulation input          & 28     & 215            & 0      & 13              & 256            & 510             \\ \hline
Step-1                    & 0      & 40 (18.6\%)    & -      & 7 (53.8\%)      & 47 (18.3\%)    & 58 (11.3\%)     \\ \hline
Step-2                    & -      & 2 (5\%)     & -      & 1 (15.2\%)      & 3 (6.4\%)    & 8 (13.8\%)    \\ \hline
Step-3                    & -      & -              & -      & 1 (100\%)      & -              & -               \\ \hline
\end{tabular}
\end{table*}

\begin{center}
{\textbf{ \Large \uppercase{Figure Legends/Captions}} }
\end{center}
\noindent

\newpage

\appendix
\setcounter{figure}{0} 
\renewcommand{\thefigure}{Supple.~\arabic{figure}} 
\setcounter{table}{0} 
\renewcommand{\thetable}{Supple.~\arabic{table}} 
\section{Validating the image reconstruction of the host galaxy}
In Supplementary Figure~\ref{fig:Host_Sim_Rec}, we present the host galaxy reconstruction for the event ID-$35$. 
The BBH parameters of this event are listed in Supplementary Table. \ref{ta:BBHParm2}. 
The host galaxy is assigned the most probable SFR. 
The first panel (first and second rows) displays results from CSST (Sloan $r$-band $600$s exposure). This exposure time is the nominal value for the CSST main survey. The second panel (third and fourth rows) shows the results obtained through a dedicated follow-up observation using JWST (F$200$W/$2\mu m$ band $1000$s exposure). 
For each panel, the first row from left to right includes: the observed image, the reconstructed image, and the normalized residuals. 
The second row for each panel from left to right includes: the reconstructed source light, the convergence, and magnification map.

\section{Validating the strong lensing model influences on time-delay reconstruction}
In Supplementary Figure~\ref{fig:TD_corner}, 
we present the posterior distribution of the time-delay from GGSL reconstruction for the most likely host galaxy. 
Here, we also tested the robustness of the time-delay reconstruction by using different lens modeling. 
We generate the time-delay and lensing image with singular isothermal ellipsoid (SIE) model.
The red and blue plots represent the time-delay reconstruction via SIE and ellipsoid power law (EPL) lens models, respectively.
We can conclude that different lens modellings do not have a strong impact on the time-delay reconstruction. Physically, this occurs because the spatial distribution of the lensing potential is much smoother than that of the convergence field. Hence, the time-delay, which reflects the lensing potential, is primarily determined by the total mass within the Einstein radius and is insensitive to the internal mass distribution. Drawing from the experience with lensed quasars, different mass models can fit the data almost equally well, yielding very similar total mass estimates.

\section{Validating the population properties of the microlensing identifiable events}
Supplementary Figure~\ref{fig:statistic}
illustrates the distribution of SLGW properties, including the logarithmic values of GW SNR $\rho$, absolute magnification $|\mu|$, and microlensing convergence $\kappa_*$. 
The red and blue plots represent microlensing identifiable events (number is $58$) and all lensing events (number is $510$) detected by CE with SNR $> 12$, respectively. 
It is noticeable that the microlensing identifiable events do not necessarily require an extremely strong macro magnification.
This phenomenon is different from the geometric microlensing, which necessitates high strong lensing magnification to enlarge the area of microlensing caustics, ensuring an high microlensing magnification rate when the source crosses the caustics. 
In wave optics, the scale at which interference occurs is typically several hundred times larger than that of a single microlensing caustics.
As seen from Eq.~($2$) in the main text, to generate a longer microlensing time-delay, this expression does not require high macro magnification, as long as some massive microlenses are located far away from the macro image center, namely large $\boldsymbol{x}^{i}$ value in the $t_\text{m}(\boldsymbol{x},\boldsymbol{x}^{i})$ term. These microlenses can produce microlensing images in the far field with large time-delays, which play a crucial role in producing the interference pattern.
It is worth noting that there are some subtleties in choosing the diffraction integral area to ensure the numerical convergence of the fast oscillation issue. 
The numerical recipe adopted in this work can safely ensure the convergence~\cite{Shan:2022xfx}. To support our arguments, Supplementary Figure \ref{fig:by_chance} presents the posterior distributions of the source redshift, the logarithmic absolute value of macro magnification, and the logarithmic value of $\kappa_*$ for all observed quadruple-image SLGWs over a 30-year period (represented by the grey shaded region, with a total of $91$ events), as well as for event ID-35 (indicated by the red star). It is evident that event ID-35 is not an outlier. Notably, its macro magnification is less than unity.

\section{Validating the effectiveness of \texttt{cWB} in identifying microlensing imprints}
In this section, we tested the necessity of using \texttt{cWB} in identifying SLGW. 
To do this, we calculated the match between the raw data (not processed by \texttt{cWB}) and the reconstruction results from \texttt{Bilby}. 
The findings, presented in Supplementary Figure~\ref{fig:match_noisy}, show a significant decrease in the match for both the unlensed and lensed events. 
In this scenario, only two events could be identified. 
Therefore, employing template-independent methods, such as \texttt{cWB}, to reconstruct the signal is crucial, as it significantly enhances the ability to identify microlensing events.
Apart from the \texttt{cWB} method, there are other template-independent approaches for reconstructing gravitational waveforms, such as the BayesWave~\citep{Cornish_2015} method and the nonorthogonal wavelet transformation method proposed by Roy~\citep{PhysRevResearch.4.033078}, among others.
In the future, it will be valuable to test the capability of different methods for reconstructing microlensing imprint. 
By searching for more effective template-independent waveform reconstruction methods, one can identify more SLGW events.

\section{Validating the origin of the interference imprint}
Supplementary Figure~\ref{fig:micro_image}
provides a detailed illustration of the wave optics effect. 
The left panel presents the time domain magnification factor.
The blue curve depicts the diffraction integral result, while red stars indicate the geometric images obtained using the ray-tracing method. 
It can be seen that some geometric images exhibit time-delays $>10^{-2}\mathrm{s}$. 
The corresponding interference fringes on the GW waveform appear in the range around $100 \mathrm{Hz}$, which is the most sensitive frequency range for ground based GW observatory. It makes that we can utilise microlensing induced signature to pinpoint the SLGW event.  
The right two panels illustrate the frequency domain magnification factor. 
The top panel displays the absolute value, whereas the bottom panel shows the complex phase. 
The blue and red curves represent the full diffraction integral and geometric optics limit results, respectively.
The geometric optics approximation can provide the amplitudes that are relatively consistent with the full diffraction integral. However, there are noticeable differences when it comes to the phase factor. To generate an accurate microlensing signal, we use the full diffraction integral in our simulation.

\section{Validating the detection significance of the identifiable microlensing events}
It is highly unlikely that random noise could produce significant waveform mismatches leading to a false positive detection of SLGWs. A rigorous assessment of this risk would involve calculating the false alarm probability for SLGW identification. However, due to the computational limitations of this study—approximately \( 1.1 \times 10^5 \) events per year—providing such an estimation is not feasible within the current scope. 
Instead, we support this claim by two steps. First, 
we validated that the boundaries of the shaded region in Figure $2$ of the main text remain unchanged when comparing the estimation results for both 100 and 200 GWs. Second, we calculated the metric distance between the match values of the identifiable events and the match value at the envelope boundary, as illustrated in Supplementary Figure \ref{fig:match_evidence}.
The envelope boundary is determined by the lowest match value among the false positives. As shown in Extended Data Fig. 2, this boundary remains unchanged for SNR values greater than $50$, even after doubling the false positive samplings and accounting for spin precession.
Here, we use the standard deviation of the identifiable events alone to measure the distance from the boundary.

In Supplementary Figure \ref{fig:match_evidence}, each point represents an event outside the light grey shaded area with SNR values greater than $50$, as shown in Figure 2 of the main text.
The $x$-axis and $y$-axis denote the GW SNR and metric distance, respectively. The horizontal dashed line marks the $5\sigma$ threshold. 
From this figure, it is clear that 85\% of the identified events show evidence exceeding the \( 5\sigma \) threshold. Notably, among events with an SNR greater than 200, only one out of 18 events shows evidence below the \( 5\sigma \) threshold.

\section{Validating the microlensing bias on SLGW parameter estimation}
In Supplementary Figure \ref{fig:bias}, we illustrate the parameter bias for all the selected single-image SLGW events from Figure 2 of the main text and their multiple-image counterparts. 
To quantify the bias introduced by the microlensing effect, we define the bias level using the following equation
\begin{equation}
\label{eq:rel_bias}
\text{bias}_{ml} = \frac{|\bar{x}_{\text {micro }}-x_{\text {inject }}|}{\sqrt{\sigma^2\left(x_{\text {micro }}\right)}} \\,
\end{equation}
where $\bar{x}_{\text {micro}}$ and $\sqrt{\sigma^2\left(x_{\text {micro }}\right)}$ represent the mean value and standard deviation of the parameter posterior distribution for SLGW with microlensing effect, respectively. 
$x_{\text{inject}}$ is the injected true value without microlensing.
$x$ represents the GW parameters listed in the figure.
The $y$-axis represents the cumulative probability distribution function of the quantity defined in Eq.~(\ref{eq:rel_bias}). 
The $x$-axis denotes the bias level.
Three dashed vertical curves with different line widths correspond to $1~\sigma$, $2~\sigma$ and $3~\sigma$ bias level.
The first and second columns of the figure display the results for intrinsic parameters, including $q$, $\mathcal{M}$.
The third and fourth columns show the results for localization parameters, RA and DEC.
One can see that for $q$ and $\mathcal{M}$, there are more than $50\%$ events out side of $3\sigma$ interval. 
However, for RA and DEC, these values is only about $10\%$. 
The similar conclusion can also be found in Mishra et al.~\citet{Mishra:2023ddt} and Shan et al.~\citet{Shan:2023qvd}.

\begin{table*}[hp]
  \centering
  \caption{\label{ta:BBHParm2} Binary black hole parameters for ID-$35$.
  $q$ is the mass ratio, $\mathcal{M}_\mathrm{obs}$ is the chirp mass, $z_s$ is the redshift of the source, $a_1$ is the spin magnitude of the primary black hole, $a_2$ is the spin magnitude of the secondary black hole, $\theta_{jn}$ is the inclination, ra is the right ascension, dec is the declination, $\Psi$ is the polarization angle, and SNR$_1$, SNR$_2$, SNR$_3$, and SNR$_4$ represent the signal to noise ratios for the first, second, third, and fourth image in this strong lensing system, respectively.} 
   \begin{tabular}{ccccccccccccc} 
    \hline
    $q$ & $\mathcal{M}_\mathrm{obs}$ & $z_s$ & $a_1$ & $a_2$ & $\theta_{jn}$ & ra & dec & $\Psi$ & SNR$_1$ & SNR$_2$ & SNR$_3$ & SNR$_4$ \\ 
    \hline
    $0.89$ & $58.9$ & $1.66$ & $0.3$ & $0.92$ & $0.82$ & $3$ & $0.62$ & $2.3$ & $557.7$ & $629.6$ & $579.4$ & $210.5$ \\
    
    \hline
  \end{tabular}
\end{table*}

\begin{figure}
\vspace{0.2em}\centering\includegraphics[width=0.9\columnwidth]{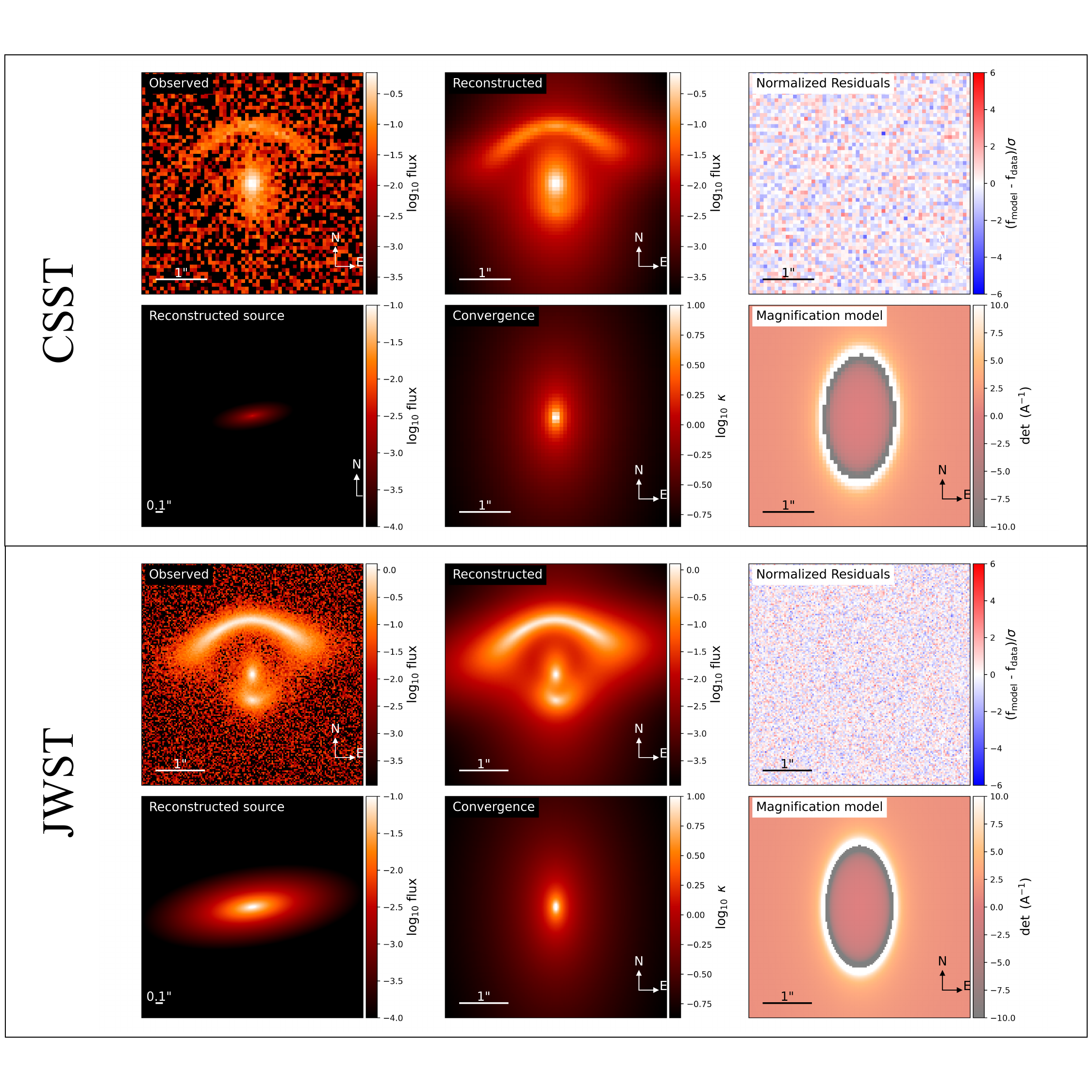}
\caption{{\bf Host galaxy reconstruction}. 
This figure illustrates the GGSL reconstruction for the most likely host galaxy of the SLGW.
The first panel (first and second rows) displays the result of CSST, while the second panel (third and fourth rows) shows the results obtained using JWST.
For each panel, the first row from left to right includes: The observed image, the reconstructed image and the normalized residuals. 
The second row for each panel from left to right includes: The reconstructed source light, the convergence and magnification map.
}
\label{fig:Host_Sim_Rec}
\end{figure}

\begin{figure}
\vspace{0.2em}\centering\includegraphics[width=0.7\columnwidth]{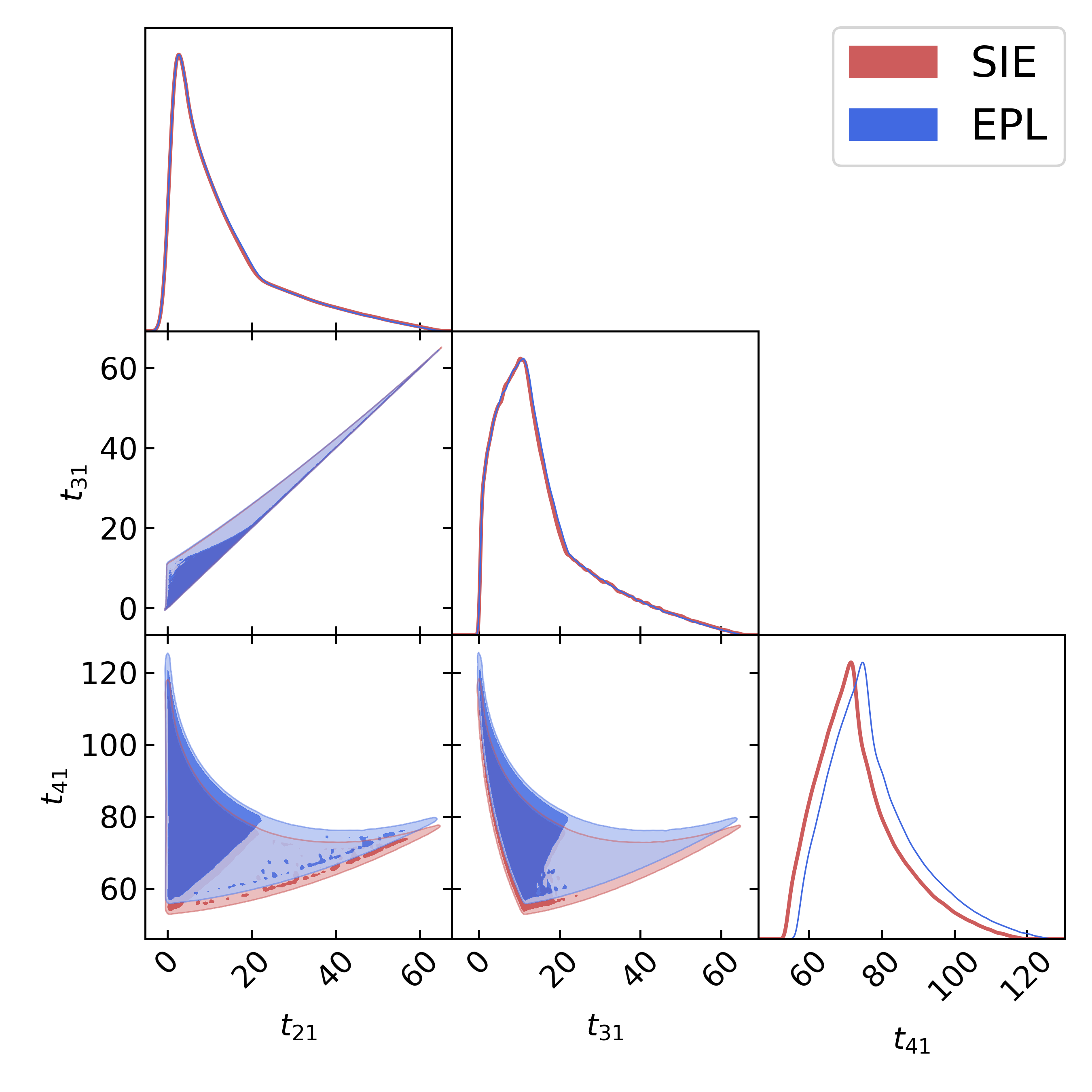}
\caption{{\bf Posterior distribution of time-delay for different lens model}. 
This figure illustrates the results of the time-delay reconstruction in the quadruple region for the most probable host galaxy of the SLGW.
The input data is generated via an SIE model.
The red and blue plots represent the reconstruction results obtained using the SIE and EPL lens models, respectively.
}
\label{fig:TD_corner}
\end{figure}

\begin{figure}
\vspace{0.2em}\centering\includegraphics[width=0.7\columnwidth]{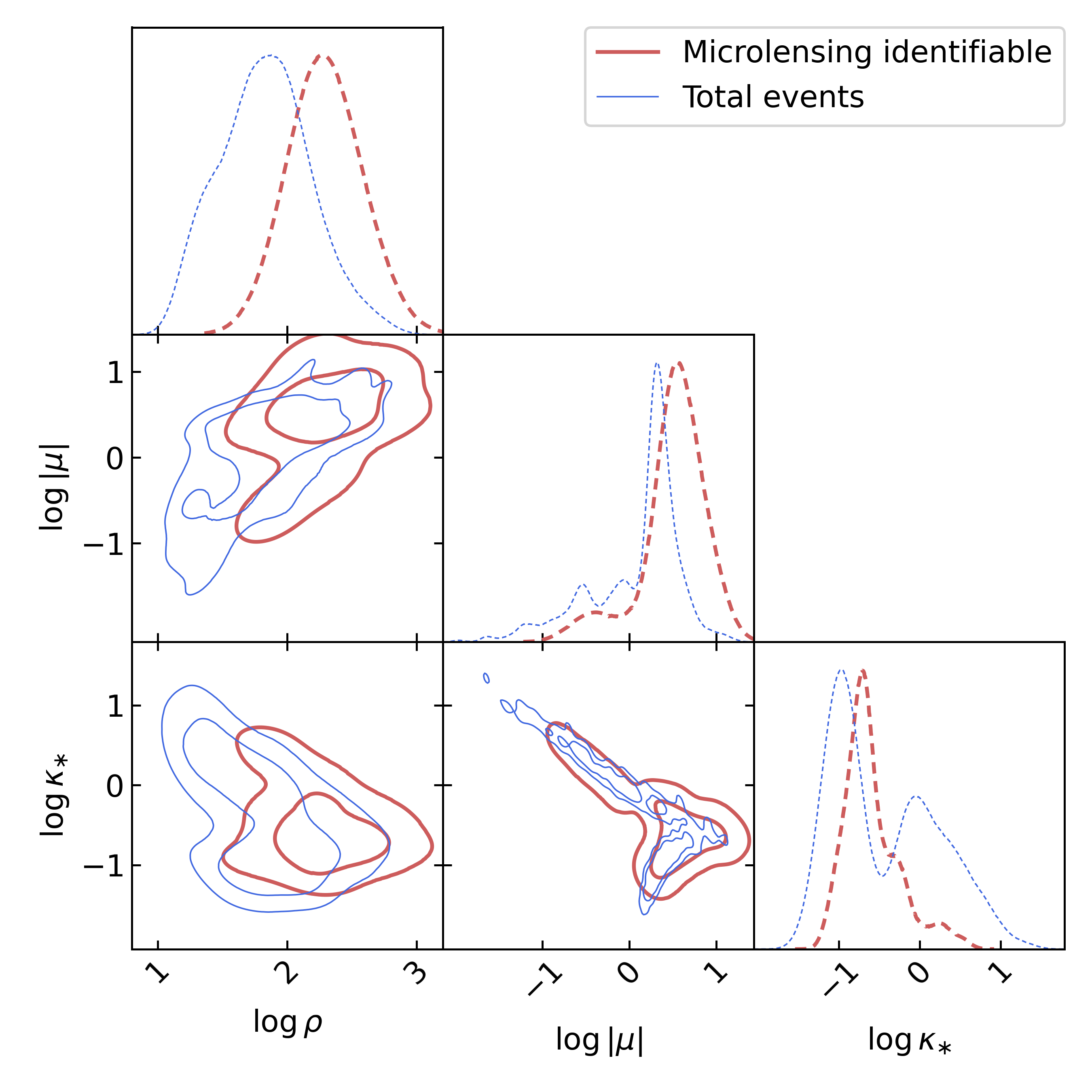}
\caption{{\bf Statistics of the identifiable events vs. all lensing events}. 
This figure illustrates the distribution of SLGW properties, encompassing the logarithmic values of GW SNR $\rho$, absolute magnification $|\mu|$, and  microlensing convergence $\kappa_*$.
The red and blue plots represent microlensing identifiable events and all lensing events, respectively.
}
\label{fig:statistic}
\end{figure}

\begin{figure}
\vspace{0.2em}\centering\includegraphics[width=0.7\columnwidth]{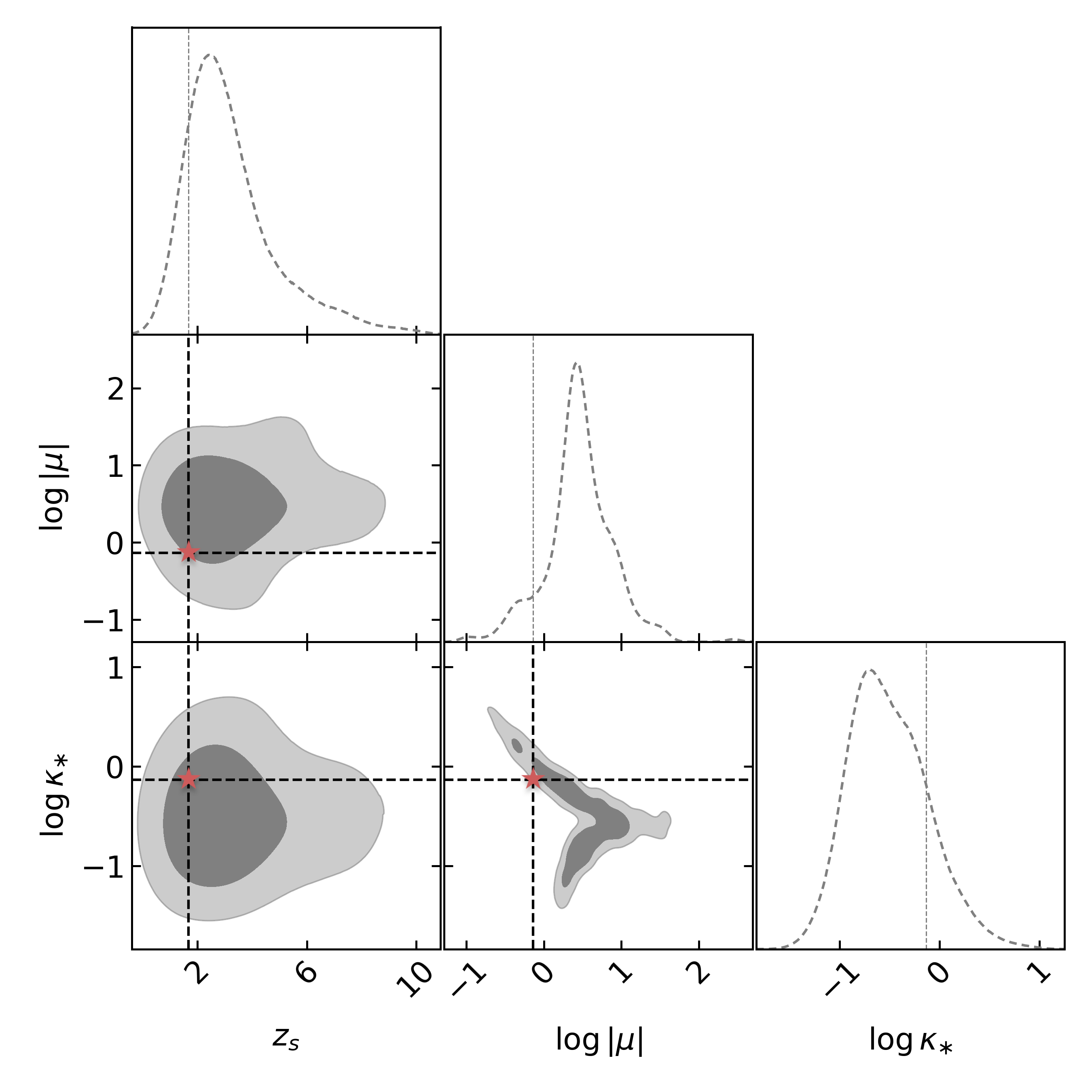}
\caption{{\bf Statistics of the identified event}. 
This figure shows the distribution of source redshift $z_s$, logarithmic absolute value of macro magnification $\log|\mu|$, and logarithmic value of microlensing convergence $\log\kappa_*$ for all the observed quadruple-image SLGWs over $30$ years observation runs with $80\%$ duty circle.
The red star stands for the event ID-35 described in the main text.
}
\label{fig:by_chance}
\end{figure}

\begin{figure}
\vspace{0.2em}\centering\includegraphics[width=\columnwidth]{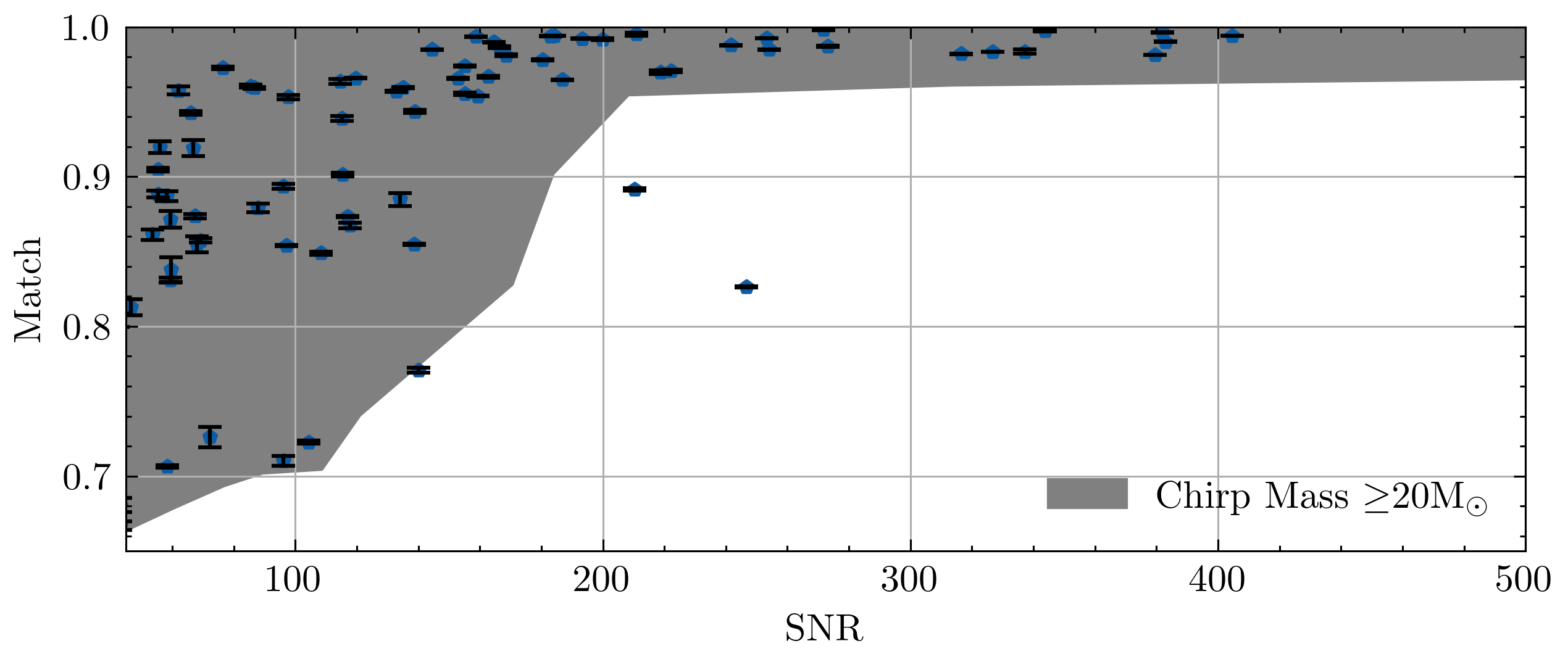}
\caption{{\bf False alarm probability test of noisy data}.
This figure shows the match between the raw data (signal + noise) and \texttt{Bilby}’s posterior results as a function of GW SNR. The grey shaded areas represent the envelope of the lowest match values for all unlensed events. The blue pentagrams (indicating the mean value) with black error bars (representing the $90\%$ confidence interval) depict the match results for SLGWs.
This plot uses a cut-off at a detector-frame chirp mass of $\mathcal{M}z \ge 20 M{\odot}$. }
\label{fig:match_noisy}
\end{figure}

\begin{figure}
\vspace{0.2em}\centering\includegraphics[width=\columnwidth]{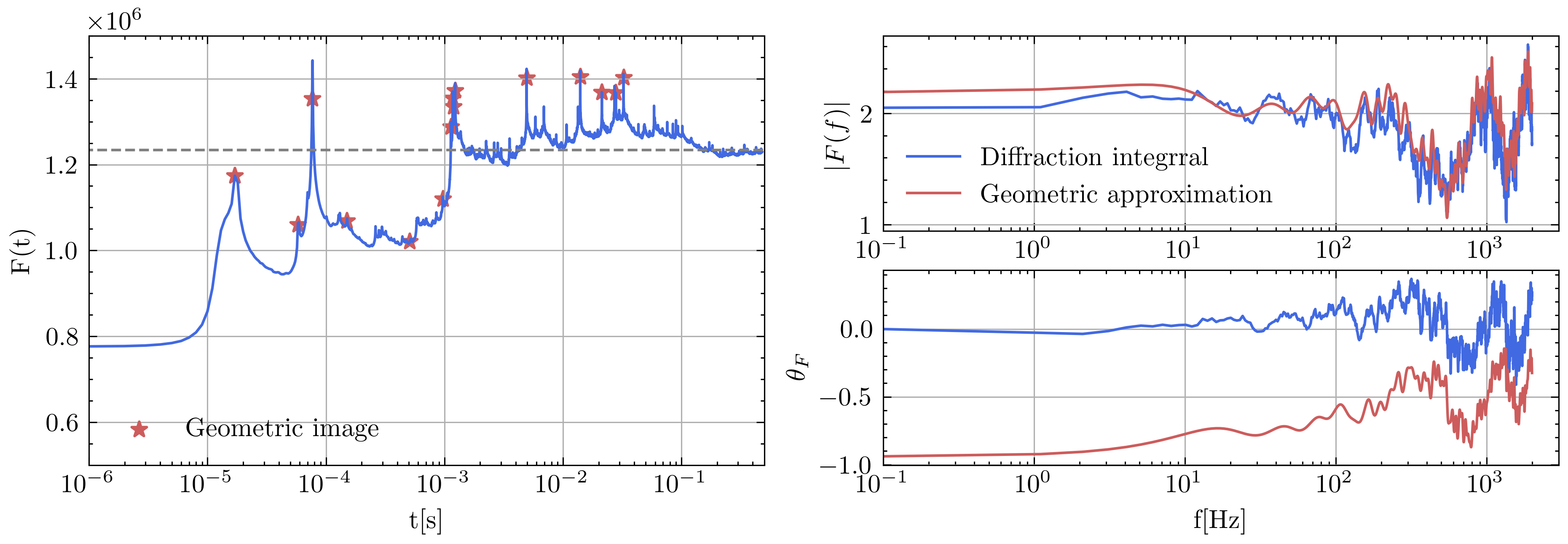}
\caption{{\bf Diffraction integral and geometric approximation result}. 
The left panel displays the results of the magnification factor in the time domain.
The blue curve represents the result obtained from the diffraction integral, while the red stars depict the geometric images found using the ray-tracing method.
The right two panels depict the magnification factor in the frequency domain.
The top panel illustrates the amplitude, and the bottom panel shows the phase.
The blue curve represents the results obtained from the full diffraction integral, while the red curve depicts the results from the sum of geometrical optics images (geometric optics limit). 
}
\label{fig:micro_image}
\end{figure}

\begin{figure}
\vspace{0.2em}\centering\includegraphics[width=0.75\columnwidth]{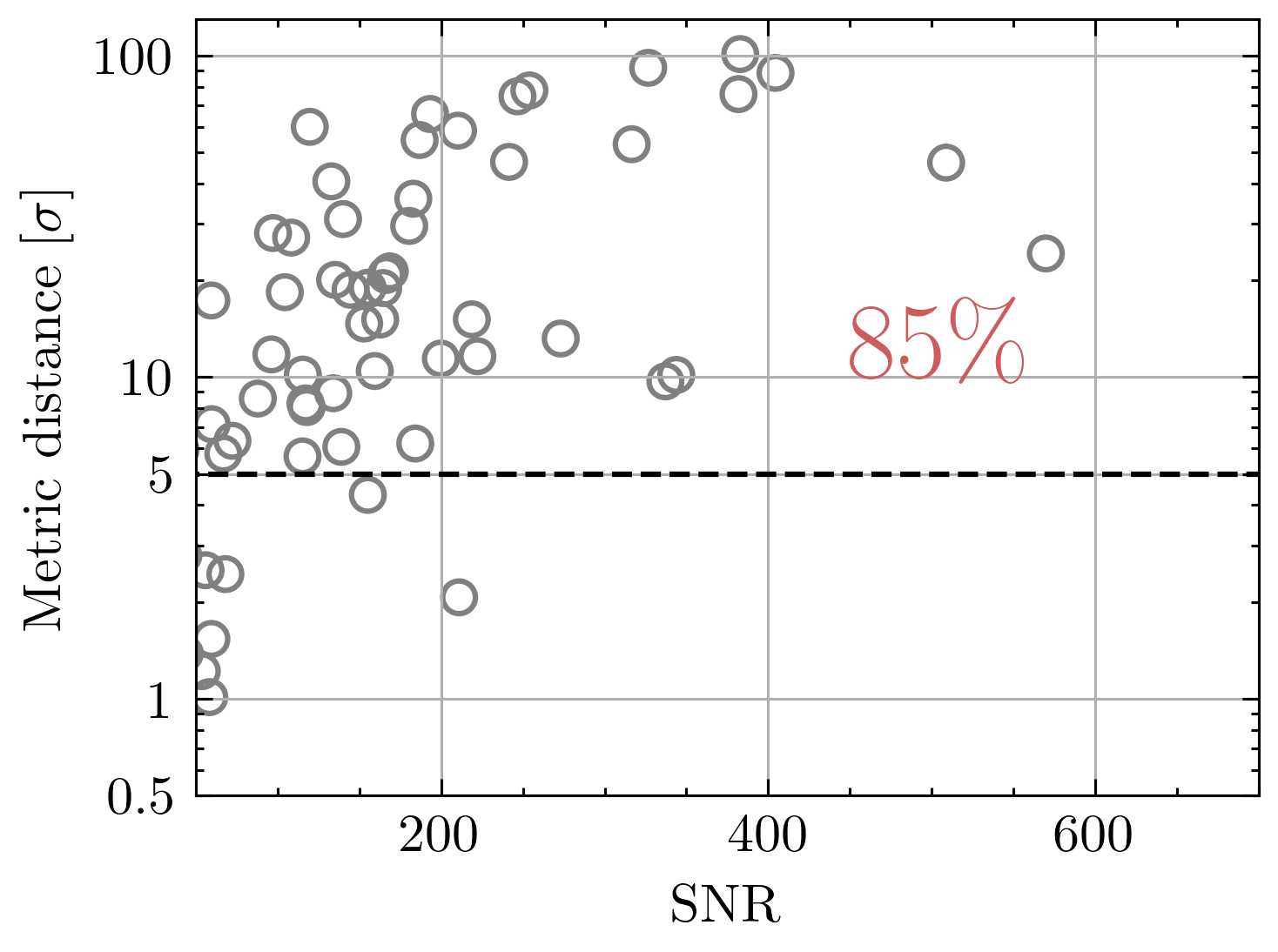}
\caption{{\bf Microlensing identification significance}.
This figure shows the metric distance between the match values for the identifiable events and the envelope boundary in Figure 2 of the main text. Each point represents an event outside of the light grey shaded area in Figure 2 of the main text. 
The distance of each events are measured by their own standard deviations. 
The $x$-axis denotes the GW SNR greater than $50$, and $y$-axis indicates the detection significance.
The horizontal dashed line marks the $5\sigma$ threshold. From this figure, it can be observed that the detection significance for $85\%$ of the identified events exceeds $5\sigma$.
}
\label{fig:match_evidence}
\end{figure}

\begin{figure}
\vspace{0.2em}\centering\includegraphics[width=0.9\columnwidth]{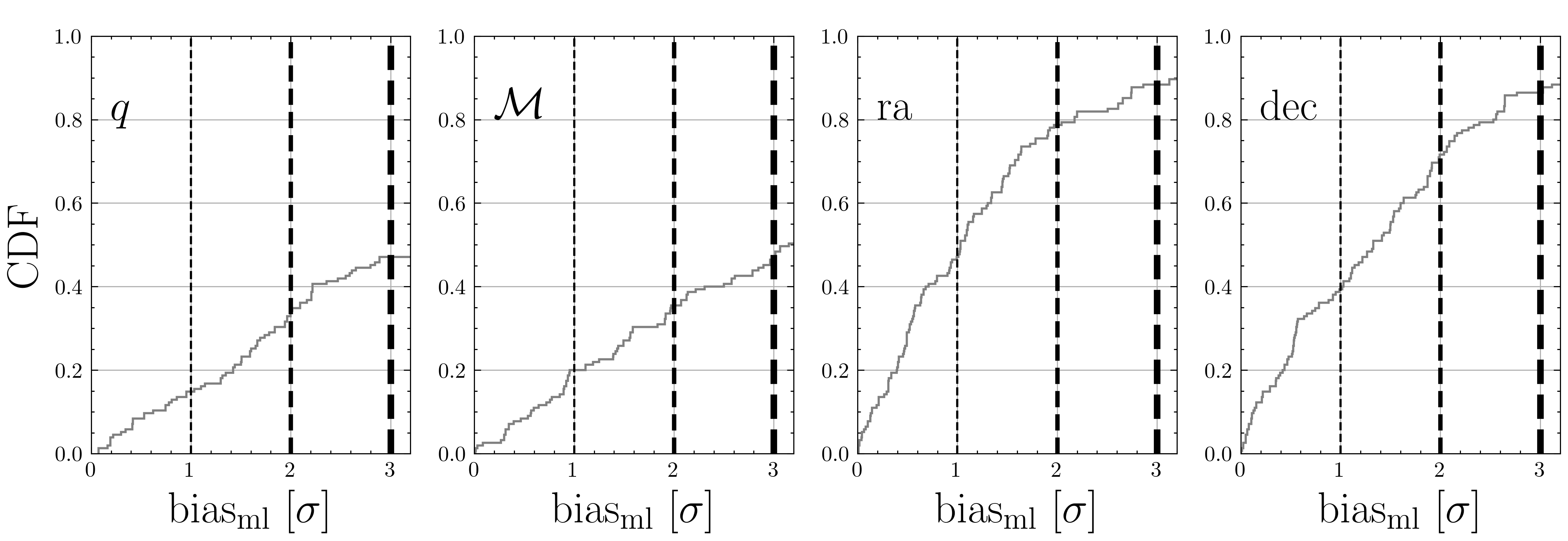}
\caption{{\bf SLGW parameter bias}. 
This figure shows the parameter bias for intrinsic parameters: mass ratio $q$ and chirp mass $\mathcal{M}$, and sky localization parameters: right ascension (ra) and declination (dec).
The $y$-axis is the cumulative probability distribution function (CDF) of the quantity defined in Main Equation (8). 
The $x$-axis denotes the bias value.
Three dashed vertical curves with different line widths correspond to $1~\sigma$, $2~\sigma$ and $3~\sigma$ bias level.
}
\label{fig:bias}
\end{figure}

\begin{figure}
\vspace{0.2em}\centering\includegraphics[width=0.7\columnwidth]{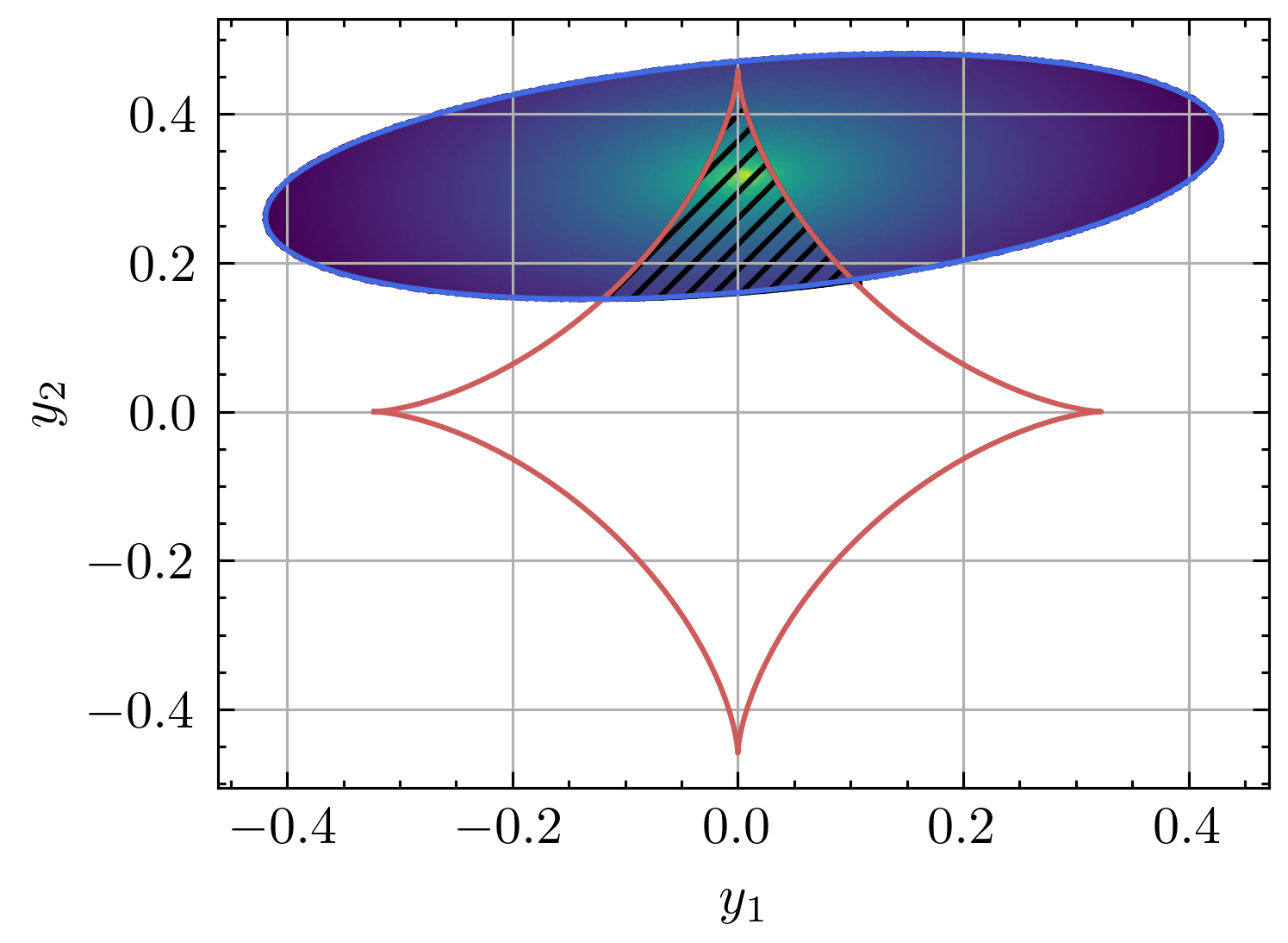}
\caption{{\bf Caustic curve and galaxy light distribution}. 
The red curve represents the caustic of a lens galaxy, while the elliptical region indicates the half-light radius of a source galaxy, with the color (from blue to yellow) representing the source light flux (from weak to strong) distribution. 
The shaded region represents the quadruple image region in the source galaxy.
The $y_1$ and $y_2$ axes represent the coordinates normalized by the Einstein radius in the source plane.
}
\label{fig:quadruple_region}
\end{figure}

\begin{figure}
\vspace{0.2em}\centering\includegraphics[width=\columnwidth]{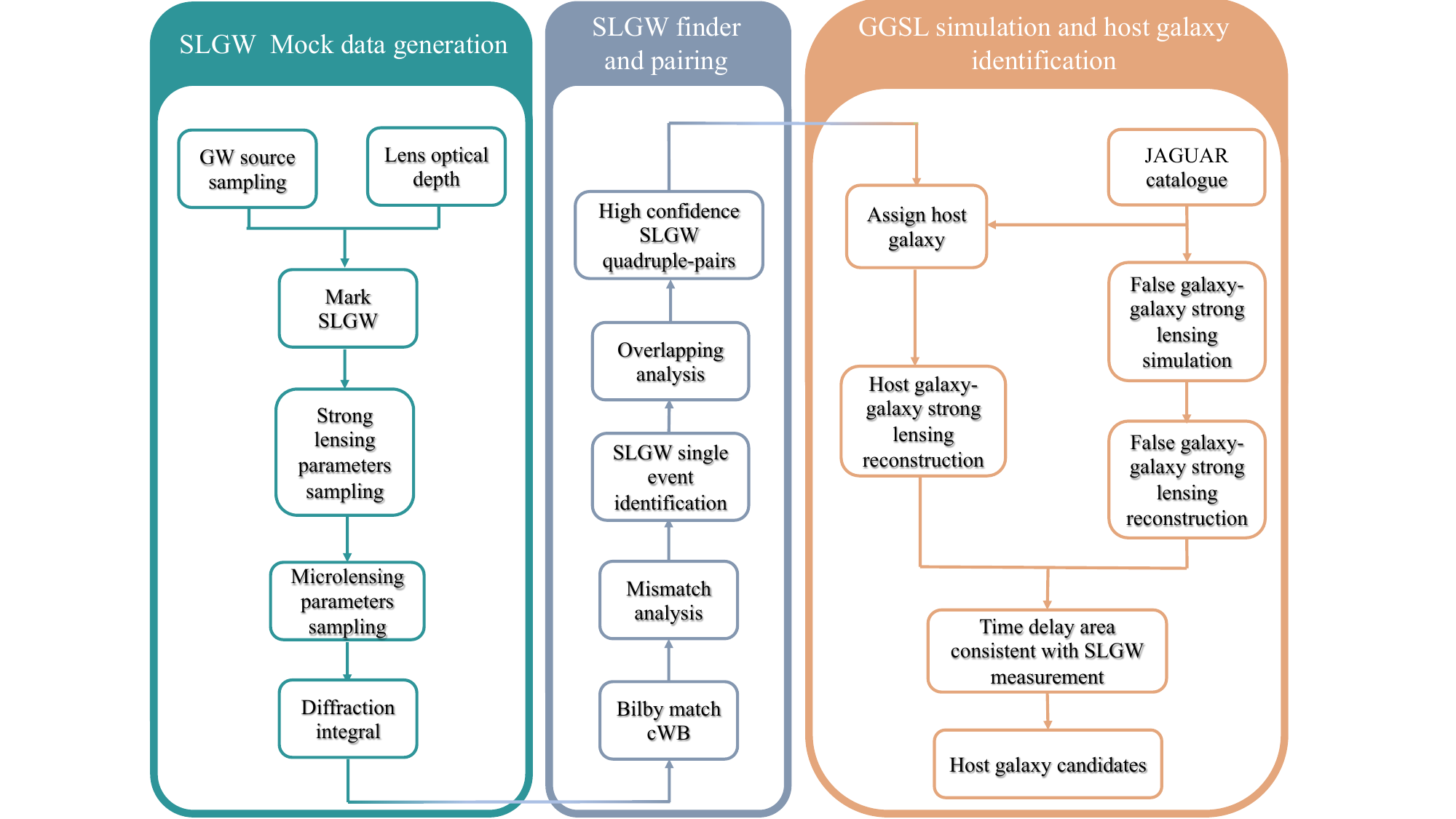}
\caption{\textbf{Flowchart of the simulation}. 
This figure shows the simulation procedures introduced in Method section.
Difference colors stand for three main simulation sections.
}
\label{fig:flowchart}
\end{figure}

\end{document}